\documentclass[prl,twocolumn,english,superscriptaddress,floatfix,nobalancelastpage,showkeys]{revtex4-2} 
\pdfoutput=1 

\usepackage{stylesheet}
\usepackage{empheq,lipsum}
\usepackage{breqn}
\usepackage{dsfont}

\renewcommand{\ell}{J}

\begin{document}

\notoc

\title{\AE\ codes}

\author{Shubham~P.~Jain}
\email{sjn@umd.edu}
\affiliation{\QUICS}

\author{Eric~R.~Hudson}
\affiliation{\CQSE}
\affiliation{\UCLA}

\author{Wesley~C.~Campbell}
\affiliation{\CQSE}
\affiliation{\UCLA}

\author{Victor~V.~Albert}

\affiliation{\QUICS}

\date{\today}

\begin{abstract}

Diatomic molecular codes [\href{https://arxiv.org/abs/1911.00099}{arXiv:1911.00099}] are designed to encode quantum information in the orientation of a diatomic molecule, allowing error correction from small torques and changes in angular momentum.
Here, we directly study noise native to atomic and molecular platforms --- spontaneous emission, stray electromagnetic fields, and Raman scattering --- {and show that diatomic molecular codes fail against this noise. 
We derive simple necessary and sufficient conditions for codes to protect against such noise.}
We also identify existing and develop new absorption-emission (\AE) codes that are more practical than molecular codes, require lower average momentum, can directly protect against photonic processes up to arbitrary order, and are applicable to a broader set of atomic and molecular systems.
\end{abstract}

\maketitle

Quantum technologies typically store, process, and transport quantum information in two states of simple quantum objects, such as atoms, photons, or electrons.
As technical capabilities increase, it becomes reasonable to consider harnessing more complicated quantum objects 
whose large state space can provide enough redundancy for an extra layer of protection against noise \eczoo[]{single_subsystem}.

Molecules provide an attractive platform for this endeavor, as their complexity is tunable from simple diatomics to large molecules like DNA and beyond, and can thus be matched to the desired task.
Further, most molecules are amenable to fast control via microwave radiation, meaning that they combine the convenient control of solid-state qubits with the desirable decoherence properties of atomic qubits.

As a result, there has been considerable experimental effort to bring molecules into the quantum toolbox.
Diatomic molecules have been prepared in single quantum states and entangled~\cite{yan_observation_2013,ni_dipolar_2018, seeselberg_extending_2018, burchesky_rotational_2021, cairncross_assembly_2021, gregory_robust_2021, holland_-demand_2022,christakis_probing_2023}; polyatomic molecules are beginning to be explored with already some success in trapping and cooling~\cite{ertmer_laser_1985, weinstein_magnetic_1998,mengel_helium_2000, campbell_magnetic_2007,steinecker_improved_2016,kozyryev_sisyphus_2017,mitra_direct_2020, stollenwerk_cooling_2020, mitra_pathway_2022,sawaoka_zeeman-sisyphus_2023}; and a number of laser-free quantum logic schemes have been proposed~\cite{mills_dipolephonon_2020, campbell_dipole-phonon_2019,mintert_ion-trap_2001, johanning_individual_2009, ospelkaus_microwave_2011, brown_single-qubit-gate_2011, timoney_quantum_2011,sutherland_versatile_2019}.

{There are only a few proposals \cite{demille_quantum_2002, zeppenfeld2023robust, wei_quantum_2016} for using the \textit{entire} molecular structure for quantum logic}. 
A recent \eczoo[molecular code]{molecular}~\cite{albert_robust_2020}  proposal revealed the possibility of utilizing various rotational state spaces for a robust encoding of information. Such codes protect information against noise that causes the angular position or momentum to change by only a small amount.
We show that such a rigid-body noise model is not always relevant to the physical noise present in more general systems, necessitating a search for other types of encodings. Molecular code states are also difficult to create, requiring an infinite superposition of angular momentum eigenstates and, ideally, a high average angular momentum.
A more compact code could thus be more amenable to current and near-term quantum devices.

\begin{figure}[t!]
    \centering
    \includegraphics[width = \columnwidth]{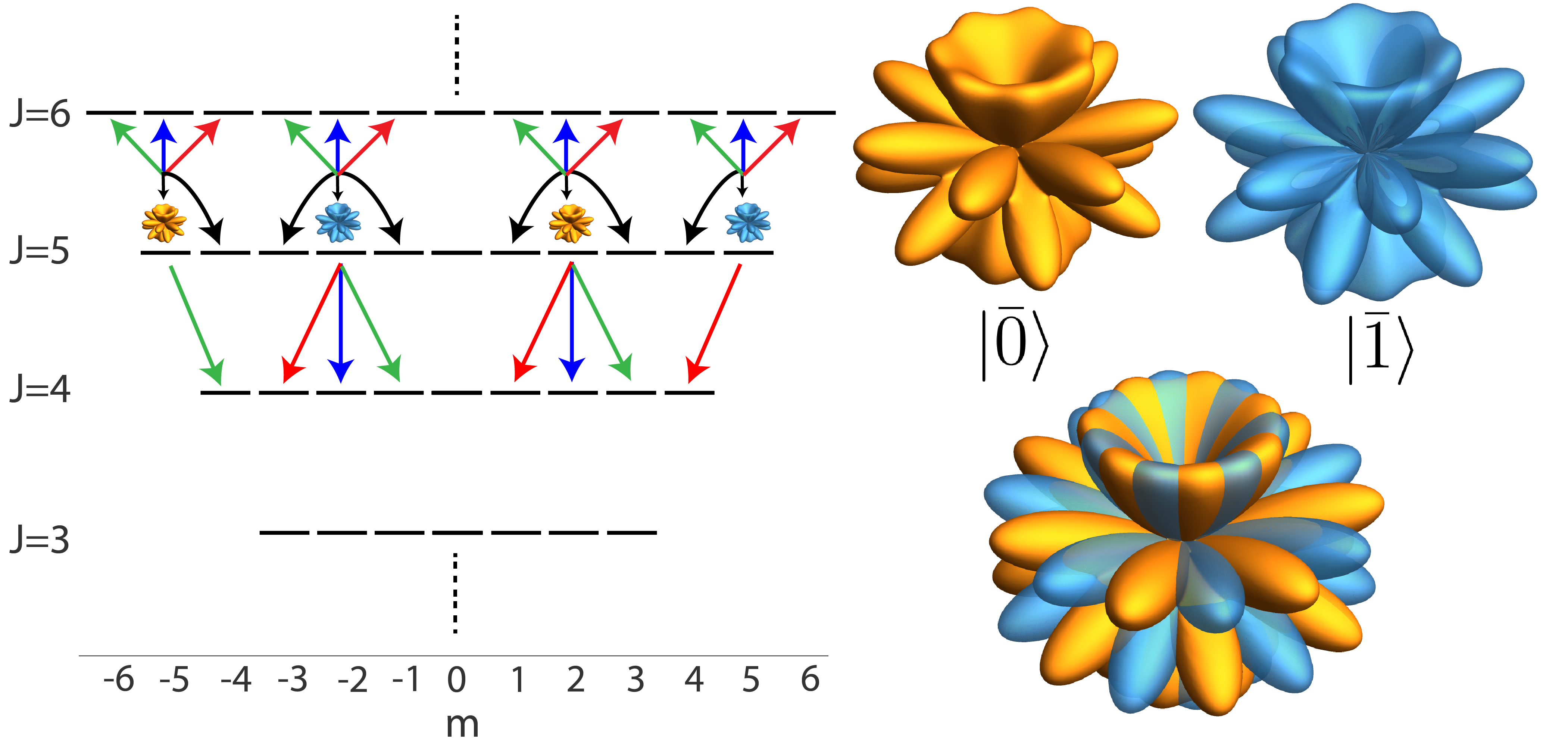}
    \caption{Representation of a $J=5$ counter-symmetric \AE\ code that protects against order-one angular-momentum transitions in the angular momentum \((J,m)\) basis (left) and as the absolute value of the spherical wavefunctions (right). 
    The wavefunctions were obtained by representing each state \(|J,m\rangle\) with the spherical harmonic \(Y^J_m(\theta,\phi)\). 
    }
    \label{fig:simple-rotor-spectrum}
\end{figure}
\begin{figure}[t!]
    \centering
    \includegraphics[width = \columnwidth]{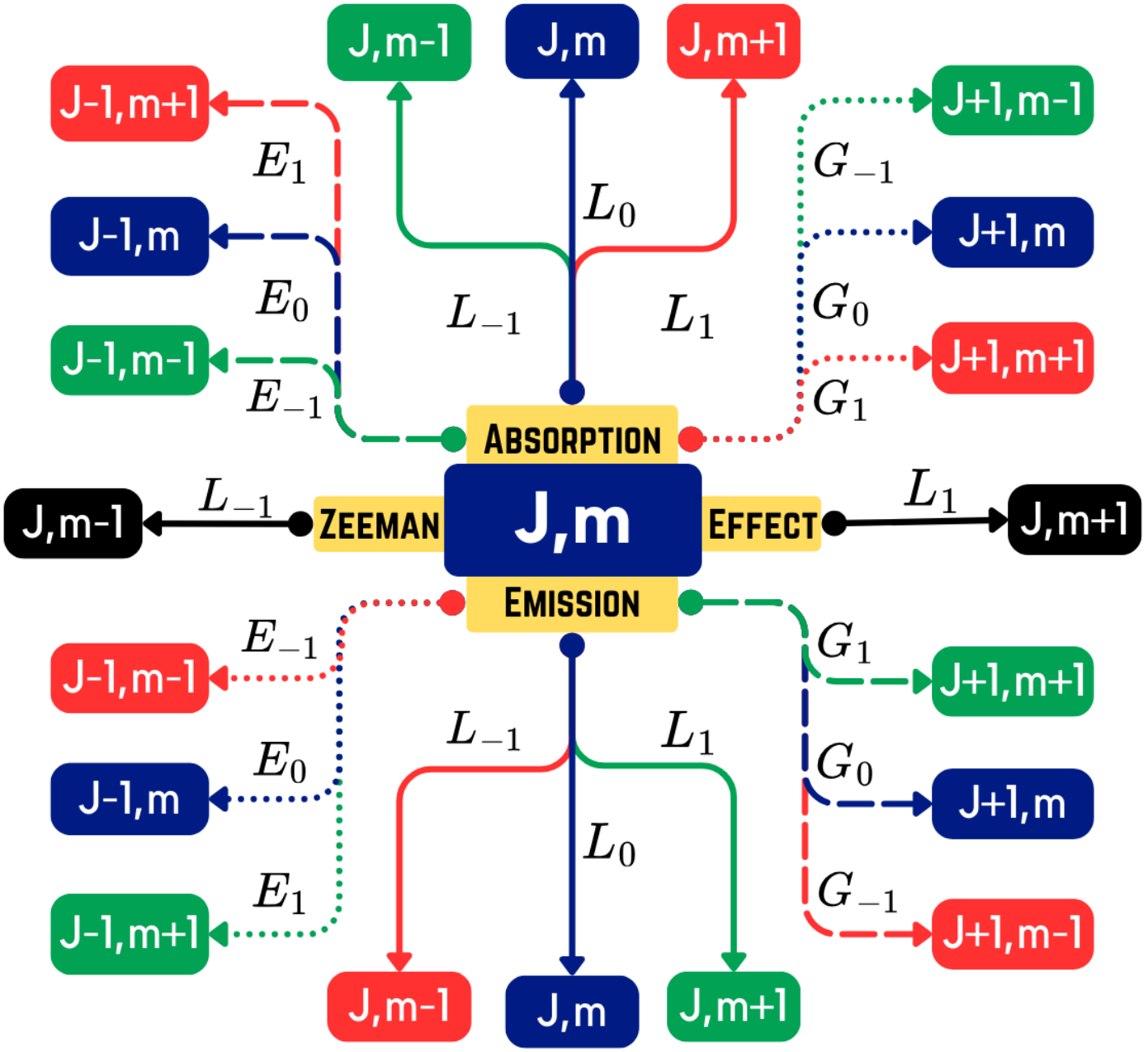}
    \caption{
    Error processes induced by absorption, emission, and the Zeeman interaction affect momentum degrees of freedom in one of nine different ways, corresponding to the nine operators $\hat{E},\hat{G}$ and $\hat{L}$ from Eq.~\eqref{eqn:error_ops}.
    The dashed (solid, dotted) lines {indicate the P (Q,R) branches~\cite{hollas2004modern} with respect to the total angular momentum $J$ i.e. $\Delta J=-1\,(0,1)$ transitions in the absorption spectrum and  $\Delta J=1\,(0,-1)$ transitions in the emission spectrum.} 
    Field polarization is indicated by color with  green (blue, red) for $\sigma^-$ ($\pi, \sigma^+$) polarization.  The Zeeman interaction on the $J$ manifold supporting the code can induce pure $\delta m =\pm1$ transitions for magnetic field $\mathbf{B} \perp \mathbf{\hat{z}}$ (black) or cause dephasing through energy shifts for $\mathbf{B} \parallel \mathbf{\hat{z}}$ (not shown).  Since $J$ is the \emph{total} angular momentum, both absorption (upper half) and emission (lower half) change $J$ by at most one in molecules with internal angular momenta.}
    \label{fig:flex_errors}
\end{figure}
Real noise in atomic and molecular systems differs from the typically studied and relatively well-behaved Pauli-type noise \eczoo{qubits_into_qubits}.
A noise process
resulting from interactions with an environment may induce one or more transitions between a system's basis states~\cite{cohen-tannoudji_atom-photon_1998, nielsen_quantum_2011, gardiner_quantum_2015, breuer_theory_2007},
corresponding to a non-unitary operator acting \textit{only} on the basis states participating in the transitions.
An important observation is that the number of transitions induced by a unique noise process is inversely correlated with a code's ability to protect against such a process.

For example, ladder-type photon loss or gain errors of a harmonic oscillator induce the maximum (infinite) number of transitions, and numerous well-performing bosonic codes can protect against such processes \eczoo{oscillators}.
On the other hand, a native anharmonicity in the system makes it possible for the environment to resolve some of these transitions into distinguishable processes, inducing noise that is uncorrectable by such codes~\cite[Sec. VIII]{albert_performance_2018}.

This would appear to be a serious limitation, as distinguishable events in ubiquitous processes such as spontaneous emission, spontaneous Raman scattering, and blackbody radiation induce transitions between only a few individual states of atoms and molecules.
{
When this noise is recast in terms of rotations and angular momentum shifts, we see that it contains terms that are not correctable by diatomic molecular codes
(see Appx. A).
However, though the energy and polarization of an absorbed or emitted photon reveal a great deal of information about the molecular state to the environment, we utilize the extra degeneracy in the \(z\)-component of angular momentum to construct new types of codes --- the \AE~(absorption-emission) codes --- that are robust to {these} common sources of error in molecular systems.

Although we primarily consider molecular systems, we find that \AE\ codes are compact enough to also be hosted in some atoms and atomic ions as long as they have a Zeeman manifold with sufficiently large total angular momentum.
Thus they provide a means to correct for spontaneous emission error in atomic laser-driven gates, which is the fundamental limit to gate fidelity in those systems ~\cite{wineland_quantum_2003,ozeri_errors_2007}.
We present the construction of \AE\ codes and discuss example encodings in molecules and atoms.

Our results apply to any system admitting multiple \(2J+1\)-dimensional irreducible representations
of total
\(SU(2)\) angular momentum \(\ell\). For a given system, this momentum may represent a combination of various nuclear, electronic, and rotational angular momenta; we use $J$ here to denote total angular momentum irrespective of the underlying structure.
The state space we consider is spanned by states \(\{|\ell,m\rangle\}\),
where \(\ell\) is a non-negative integer or half-integer, and \(m\) denotes
the \(z\)-axis projection ranging from \(-\ell\) to \(\ell\).

The goal of error correction is to pick a subspace of a system's state space that is robust to noise.
A code subspace, or codespace, can be defined by a basis of its \textit{codewords} $\{|\bar{i}\rangle\}$.
We consider codespaces of \textit{fixed} total angular momentum $J$. For the case of a logical-qubit encoding, the codewords are
\begin{equation}\label{eq:codewords}
    \ket{\bar{0}}={\textstyle \sum_{m}}\alpha_{m}\ket{\ell,m}\quad\text{and}\quad\ket{\bar{1}}={\textstyle \sum_{m}}\beta_{m}\ket{\ell,m},
\end{equation}
where \(\alpha_m,\beta_m\) are complex coefficients that ensure proper normalization.
The total-momentum restriction ensures that our encodings apply not only to rotational states of diatomic molecules, but also to sufficiently large atomic subspaces of fixed total momentum.

Dominant noise processes acting on code states should be detectable and correctable.
A given set of error processes is correctable if all pairs of representative Kraus operators~\cite{nielsen_quantum_2011} from the set $\{\hat{E}_{a}\}$ satisfy the Knill-Laflamme (KL) conditions~\cite{knill_theory_1997, bennett_mixed-state_1996}
\begin{equation} \label{eqn:knil_laflamme_conditions}
    \braket{\bar{i}|\hat{E}_a^\dagger \hat{E}^{}_b|\bar{j}}=\delta_{ij}c_{ab}\quad\quad\text{(KL conditions)}.
\end{equation}
The Kronecker delta, $\delta_{ij}$, implies that no two orthogonal states in the codespace should be mixed by the errors.
The requirement that the weight $c_{ab}$ is independent of $i$ ensures that the errors occur with the same probability for each code word and therefore do not distinguish between code states. In physical terms, this means that the qubit's information remains intact and unentangled with the environment after an error has occurred. {For a set of error processes to be detectable, it is sufficient for the representative Kraus operators to only satisfy the conditions $\braket{\bar{i}|\hat{E}^{}_a|\bar{j}}=\delta_{ij}c_{a}$}.

Molecular systems are inherently anharmonic, meaning that any arbitrarily-chosen transition is likely to be energy-resolvable from another arbitrarily chosen transition, even of the same rank. 
Distinguishable noise processes that induce transitions between only a few individual eigenstates are actually harder to correct, with processes inducing transitions between a single pair of eigenstates placing severe restrictions on the codespace.

Using the KL conditions \eqref{eqn:knil_laflamme_conditions}, we observe that a distinguishable noise process connecting single eigenstates, \(\hat{E} = |\ell^\prime,m^\prime\rangle\langle\ell,m|\) for some \(\ell^\prime \neq \ell\) or \(m^\prime\neq m\), automatically precludes the state \(|\ell,m\rangle\) from being used to construct the code.
In other words, we are required to set \(\alpha_m = \beta_m = 0\) in Eq. \eqref{eq:codewords} if there is any hope of constructing a code protecting against such a process.
This idea holds more generally, as demonstrated by the following theorem (see Appx. B for proof).

\begin{theorem} \label{thm:nogo} Given a Hilbert space with countable basis and orthogonal vectors $\ket{\psi}$ and $\ket{\phi}$,
an error channel containing the Kraus operator $\hat{E} = \ket{\phi}\bra{\psi}$ is correctable only if the codespace is not supported by $\ket{\psi}$.
\end{theorem}
If there are sufficiently many single-transition noise processes, the above theorem restricts the state space so severely that error-correction is not possible (cf.~\cite{chessa_quantum_2021,chessa_resonant_2023}).
Fortunately, molecular transitions are not fully resolved: while the environment is able to distinguish transitions between different angular momenta \(\ell,\ell^\prime\), we show that error correction is still possible if the environment does not resolve the different \(z\)-axis components \(m,m^\prime\).
This occurs when states $\ket{\ell,m}$ of a given fixed-\(\ell\) manifold are degenerate in \(m\) to the precision of the linewidth of the transition associated with the error --- a reasonable assumption for many atomic and molecular systems~\cite{foot_atomic_2005}.

A common source of error when dealing with atoms or molecules is the dipole-absorption or spontaneous dipole-emission of a photon, which leads to a change in the angular momentum and its projection by at most one. Any noise channel acting on the system can be expressed in the Kraus representation \cite{nielsen_quantum_2011}, with spontaneous emission from a fixed-$J$ subspace yielding the nine possible Kraus operators
\begin{align} 
   &\hat{E}_{0}\propto\,{\textstyle \sum_{m}\scriptstyle{\sqrt{\ell^{2}-m^{2}}}} \, \ket{\ell-1,m}\!\bra{\ell,m}\nonumber\\
   &\hat{E}_{\pm1}\propto\,{\textstyle \sum_{m}\scriptstyle{\sqrt{(\ell\mp m-1)(\ell\mp m)}}} \, \ket{\ell-1,m\pm1}\!\bra{\ell,m}\nonumber\\
   &\hat{G}_{0}\propto\,{\textstyle \sum_{m}\scriptstyle\sqrt{(\ell+1)^2-m^2}}\ket{\ell+1,m}\bra{\ell,m}\nonumber\\
   &\hat{G}_{\pm1}\propto\,{\textstyle \sum_{m}\scriptstyle\sqrt{(\ell\pm m+1)(\ell\pm m+2)}}\ket{\ell+1,m\pm1}\bra{\ell,m}\nonumber\\
   &\hat{L}_{0}\propto\,{\textstyle \sum_{m}}m\ket{\ell,m}\bra{\ell,m}\nonumber \\
   &\hat{L}_{\pm1}\propto\,{\textstyle \sum_{m}}{\scriptstyle \sqrt{(\ell\mp m)(\ell\pm m+1)}}\ket{\ell,m\pm1}\bra{\ell,m}~.\label{eqn:error_ops}  
\end{align}
The square-root coefficients in $\hat{E}_{\delta m}$, $\hat{G}_{\delta m}$ and $\hat{L}_{\delta m}$ are proportional to the Clebsch-Gordan factors $C^{J-1,m+\delta m}_{1,\delta m;J,m},C^{J+1,m+\delta m}_{1,\delta m;J,m}$ and $C^{J,m+\delta m}_{1,\delta m;J,m}$, respectively \cite{varshalovich_quantum_1988}. These operators obey symmetry relations, $\hat{E}_{\delta m}^\dagger(J+1) = \hat{G}_{-\delta m}(J)$ and $\hat{L}_{\delta m}^\dagger = \hat{L}_{-\delta m}$, where $\hat{E}(J)$ and $\hat{G}(J)$ denote the Kraus operators acting on the $J$ total angular momentum manifold.

The transitions described by these operators can change quantum numbers other than \(J\) and \(m\), but we restrict the notation to explicitly indicate only the total angular momentum labels involved, as the $m$-dependences of the amplitudes are functions of only $J$ and $m$. Various other physical processes correspond to the same Kraus operators (see Fig.~\ref{fig:flex_errors}).

Distinguishable versions of the above Kraus processes occur for \textit{each} \(\ell\).
The ability of the environment to resolve different total angular momenta renders these processes sufficiently severe so as to be uncorrectable by diatomic molecular codes \cite[Sec.~VI]{albert_robust_2020}, as can be verified by direct calculation of the KL conditions \eqref{eqn:knil_laflamme_conditions}.

Given this result and the multitude of relevant error sources, it is natural to ask: \emph{Can codewords resilient to photon absorption or emission error be found at all?}
We answer this in the affirmative by noticing that the operators \eqref{eqn:error_ops} consist of \textit{multiple} unresolvable transitions between various projection values \(m,m^\prime\).

To ensure that decays with $\delta m = \pm 1$ do not mix the codewords, we take the codestates to be superpositions of basis states separated by \textit{at least} three units in $m$, as illustrated in Fig.~\ref{fig:simple-rotor-spectrum}.
Given this ansatz, a code has only to satisfy the ``diagonal'' KL conditions,
\begin{equation}\label{eq:diagonals}
    \bra{\bar{0}}\hat{K}_{\delta m}^\dagger \hat{K}^{}_{\delta m}\ket{\bar{0}}=\bra{\bar{1}}\hat{K}_{\delta m}^\dagger \hat{K}^{}_{\delta m}\ket{\bar{1}}~,~ \delta m \in\{0,\pm1\}~,
\end{equation}
where $\hat{K} \in \{\hat{E},\hat{G},\hat{L}\}$, in order to correct all the mentioned errors.

We can see that all KL matrix elements of the form $\bra{\ell,m}\hat{K}_{\delta m}^\dagger \hat{K}^{}_{\delta m}\ket{\ell,m}$ are \textit{quadratic polynomials} in $m$. As a result, these conditions are satisfied if the codes protect against the dephasing error set $\{\mathds{1},\hat{m}\}$, where \(\mathds{1}\) is the identity, and
\begin{equation}\label{eq:dephasing}
    \hat m = {\textstyle \sum_{\ell,m}} |\ell,m\rangle m \langle\ell,m|
\end{equation}
is the dephasing operator.
This is because the diagonal KL matrix elements for the dephasing error set are monomials in $m$ up to order $2$, which in turn form the diagonal KL matrix elements of the Kraus operators in Eq.~\eqref{eqn:error_ops} via linear combination.
Thus, we can recast the complicated error operators of Eq.~\eqref{eqn:error_ops} in terms of dephasing --- a smaller and simpler set of correctable noise operators.

In summary, any codewords that are sufficiently spaced in \(m\) and satisfy Eq. \eqref{eq:diagonals} for dephasing errors \textit{automatically} protect against the much larger and seemingly more complex error set
\(\{\hat{E}_{\delta m}, \hat G_{\delta m}, \hat L_{\delta m}, \hat m\}\) with \(|\delta m|\leq 1\).

\paragraph{Example codewords.}
The task of finding the codewords which satisfy the two diagonal KL constraints corresponding to the dephasing error set $\{\mathds{1},\hat{m}\}$ can be formulated as finding solutions to a simple linear system. This system can be under-constrained, and we list two families of solutions below (see Appx. C for details and more examples).

For integer $\ell$, restricting to codewords that have amplitudes which are symmetric about $m=0$, we obtain the \textit{symmetric codes}
\begin{align}
\ket{\bar{0}}&={\textstyle \sqrt{\frac{1}{2}}}\left(\ket{\ell,-m_{1}}+\ket{\ell,m_{1}}\right)\label{eqn:DL1_{c}ode}\\\ket{\bar{1}}&={\textstyle \sqrt{(1-\frac{m_{1}^{2}}{m_{2}^{2}})}}\ket{\ell,0}+{\textstyle \sqrt{\frac{m_{1}^{2}}{2m_{2}^{2}}}}\left(\ket{\ell,-m_{2}}+\ket{\ell,m_{2}}\right),\nonumber
\end{align}
for any choice of positive integers $\ell \geq 6$, $m_1\geq 3$ and with the constraint $m_2 \geq m_1 + 3$ (see Fig.~\ref{fig:simple-rotor-spectrum}).

These are highly tunable, working even when the spacing between neighboring states is non-uniform (\(m_2 \neq 2m_1\)) or when the manifold is saturated ($m_2 = \ell$).
As a consequence of the form of the Clebsch-Gordan coefficients, the basis-state amplitudes within a codeword are independent of $\ell$ and depend only on the ratio $m_1/m_2$.

A more basic \textit{error-detecting} version of these codes,
\begin{align}
\ket{\bar{0}}={\textstyle \sqrt{\frac{1}{2}}}\left(\ket{\ell,-m}+\ket{\ell,m}\right)\quad\text{and}\quad\ket{\bar{1}}=\ket{\ell,0},\label{eqn:DL1_{c}ode_{d}etection}
\end{align}
is possible for \(\ell,m \geq 2\); this can be thought of as an analogue of the \eczoo[dual-rail code]{dual_rail}~\cite{chuang_simple_1995}, detecting a single transition error.

Restricting to codewords that are \textit{counter-symmetric} about $m = 0$, we get the family of solutions 
\begin{align} \label{eqn:counter_symmetriclvl1}
\ket{\bar{0}}&={\textstyle\sqrt{\frac{m_2}{m_1+m_2}}}\ket{\ell,-m_1}+{\textstyle\sqrt{\frac{m_1}{m_2+m_1}}}\ket{\ell,m_2}\nonumber\\
\ket{\bar{1}}&={\textstyle\sqrt{\frac{m_1}{m_1+m_2}}}\ket{\ell,-m_2}+{\textstyle\sqrt{\frac{m_2}{m_2+m_1}}}\ket{\ell,m_1}.
\end{align}
This code works for both integer and half-integer $\ell \geq 9/2$, with \(m_1 \geq 3/2\) and \(m_2\geq m_1 + 3\).
Further, like the previous code, the amplitudes are independent of $\ell$ and depend only on $m_1/m_2$. For detection, $\ell \geq 3$ with \(m_1 \geq 1\) and \(m_2\geq m_1 + 2\) suffices.

\paragraph{Higher-order transitions.}
Errors causing higher-order transitions can occur if a laser is applied to the system, inducing spontaneous Raman scattering. A key observation of this work is that the patterns described above generalize, yielding a simplified error set along with corresponding codes.

An error channel
causing up to order-$n$ transitions consists of  all operators
\begin{align}
    \hat{K}_{\delta m}^{r(\delta \ell)} \propto& \sum_{|m|\leq J_0}  \bra{\ell_0+\delta \ell,m+\delta m}\hat{Y}^r_{\delta m}\ket{\ell_0,m}\nonumber \\
    &\times \ket{\ell_0+\delta \ell,m+\delta m}\bra{\ell_0,m}~,
\label{eqn:lvlnAEchannel}
\end{align}
where $ |\delta \ell| \leq r, \, r \leq n$, and the \(\hat Y\) {operators are generalised phase shifts defined by the equation $\hat{Y}^{J}_{m}\ket{\theta,\phi}={Y}^{J}_{m} (\theta,\phi)\ket{\theta,\phi}$ involving the spherical harmonics $Y^{J}_{m}(\theta,\phi)$~\cite[Tab.~V]{albert_robust_2020}.}

This model encompasses all emission, absorption, dephasing and raising/lowering noise up to order $n$ (including all electric and magnetic $2^r$-pole transitions up to $r=n$). For first-order transitions, $r=1$ and 
$\delta\ell=-1,1,0$, correspond to $\hat{E},\hat{G},\hat{L}$, respectively.
General higher-order transitions allow for changes \(|\delta m|, |\delta\ell| \leq r\), yielding a total of \((2r+1)^2\) different errors.

Choosing a sufficiently large spacing in \(m\) automatically leads to satisfaction of the off-diagonal KL conditions for many of the noise operators.
The relevant diagonal conditions, which in general include products \(\hat{K}^{r_1\dagger} \hat{K}^{r_2}\) with \textit{different} values of \(r\), are potentially problematic since each operator contains a square-root of an \(r\)-dependent polynomial in \(m\).
Surprisingly, all relevant products turn out to contain the \textit{same} square root, thereby yielding a polynomial in \(m\) of degree \(r_1+r_2\) (see Appx. D for proofs). 

The diagonal conditions are once again reduced to the conditions for a generalized dephasing operator set, $\left\{\hat{m}^k\, | \, 0\leq k\leq n\right\}$.

One general family of \AE\ codes protecting against angular momentum change 
and dephasing errors \eqref{eqn:lvlnAEchannel} up to order $n$ admits the basis (see Appx. E for details)
\begin{equation}\label{eqn:lvln-codes}
    \ket{\bar{0}/\bar{1}}=\frac{1}{2^{n}}\sum_{\text{even /odd}~k=0}^{2n+1}{\textstyle \sqrt{\binom{2n+1}{k}}}\ket{\ell_{0},-m_{0}+k(2n+1)}~,
\end{equation}
for $J_0,m_0 \geq (2n+1)^{2}/2$.
Setting $\ell_0=(2n+1)^2/2$ transforms this family into the counter-symmetric code family, to which Eq. \eqref{eqn:counter_symmetriclvl1} belongs. This family allows for the most resource-efficient packing for correcting order-\(n\) errors. For error detection, total angular momentum \(\ell_0=(n+1)^{2}/2\) suffices, along with $m_0=(n+1)^2/2$ and spacing of \(n+1\) between supported momentum states.

With the complicated error-correction conditions reduced to a simple linear system, we observe that the above codes can be relevant to other paradigms.
Using an ansatz consisting of identical spacing between any neighboring pair of states and only allowing states with \(m\geq 0\) recovers the bosonic \eczoo[binomial codes]{binomial} \cite{michael_new_2016}, or the permutation-invariant many-qubit \eczoo[GNU codes]{gnu_permutation_invariant} \cite{ouyang_permutation-invariant_2014}. The dephasing operator \(\hat m\) is mapped to the oscillator occupation number (total qubit spin) in the former (latter) case. { \eczoo[Single-spin codes]{single_spin} \cite{gross_designing_2021,gross_hardware-efficient_2021,fan_quantum_2023, omanakuttan_spin_2023,omanakuttan_multispin_2023}\cite[Sec.~III.B]{albert_robust_2020} with sufficient spacing are examples of \AE\ codes.}  We explicitly consider transitions in angular momentum \(\ell\), and removing those reduces our noise model to that of spin codes. All of our codes thus work just as well against shift errors of a single spin (cf.~\cite{aditya_sivakumar_quantum_2021}), and our variable-shifted codes generalize the binomial codes for the oscillator.

\parg{Discussion \& conclusion}
We simplify the conditions required for an encoding to protect against transitions in angular momentum due to the gain or loss of \(n\)
quanta. This allows us to identify several code families that protect against such processes up to any \(n\).
Our codes can be hosted in a total angular momentum \(J\) space which can arise from combinations of spin, electronic, or rotational, or nuclear angular momenta.

The $n = 2$ {correction} codes allow correction of spontaneous Raman scattering error and are therefore potentially useful in atom-based quantum processing, where this error sets the fundamental gate fidelity for laser-based gates~\cite{boguslawski_raman_2023}.
They require a total angular momentum of at least $J = 25/2$, which is present in a number of transition-metal atoms with isomeric nuclear states.
For example, naturally-abundant $^{180m}$Ta$^+$ is observationally stable and has nuclear spin $I = 9$ with a ground state described as $^5F$, and therefore possesses two subspaces of total angular momentum 13, and one of 14.  
Both ${}^{143}\mathrm{Nd}^+$ and ${}^{145}\mathrm{Nd}^+$ are naturally abundant with $I=7/2$, and the metastable electronic ${}^6\mathrm{L}$ state can support $n=2$ correction codes. 
Similarly, isomeric nuclear states in $^{178}$Hf$^+$ and $^{192}$Ir$^+$ with radioactive half lives of approximately 31~y and 240~y, respectively, could host the code.
Direct laser cooling of these ions has yet to be demonstrated.

If only $n = 2$ error detection is required, then atomic states with momentum $\geq 9/2$ are sufficient.
This requirement is much more easily satisfied; for example, $^{173}$Yb$^+$, which has been trapped and laser cooled in the group of one of the authors (WCC), has nuclear spin $I = 5/2$, leading to several metastable states in the ${}^2\mathrm{F}^o_{7/2}$ manifold that can host the error-detection code.
Further, In, In$^+$ and Lu$^+$ have been trapped and laser cooled~\cite{burt_demonstration_1995,kim_efficient_2009, arnold_blackbody_2018, yu_magneto-optical_2022} and possess isotopes with $I \geq 9/2$.
In the case of In$^+$ and Lu$^+$, the ground states are described as $^1$S$_0$.
Such systems could be interesting as the qubit is hosted solely in the nuclear spin and therefore likely more robust than an electronic qubit due to the smaller size of the nucleus. {Laser-cooled neutral atom species including $\mathrm{Ho}$ \cite{miao_magneto-optical_2014}, $\mathrm{Dy}$ \cite{lu_trapping_2010}, and $\mathrm{Er}$ \cite{mcclelland_laser_2006} can likewise support $n=2$ error detection codes in their ground electronic states.}

The same codes also correct $n=1$ errors, which are relevant for protection against
first-order coupling to electromagnetic vacuum and blackbody radiation and are a primary noise source for molecular qubits.
States with sufficient $J$ to host the code are available in essentially \emph{every} molecule.
For example, in simple diatomic molecules like CaF, which have recently been entangled \cite{holland_-demand_2022,bao_dipolar_2022} and $^{29}$SiO$^+$, which is promising for molecular-ion quantum logic \cite{stollenwerk_cooling_2020, zhu_high-resolution_2022}, the code can be hosted in the fourth rotationally excited state.
Similarly, our symmetric code can be hosted in the third rotatonally excited state of H$^{35}$Cl$^+$, where it may be possible to perform error correction via electric-field gradient gates at $1.6$~GHz~\cite{hudson_laserless_2021}. {Achieving this would require single-quantum-state preparation, control, and readout of the molecular ion, likely facilitated by a co-trapped atomic ion as has been demonstrated for CaH$^+$ \cite{chou_preparation_2017}. Implementation strategies using co-trapped atomic ions for general \AE\ codes have recently been proposed, and their performance has been evaluated for the counter-symmetric code~\cite{furey2024strategies}.} 

Finally, the \(n=1\) error-detecting code in Eq.~\eqref{eqn:DL1_{c}ode_{d}etection} fits into any subspace with total momentum two. Assuming that control of atomic, molecular, and color center qubits improves, this code could serve as a atomic/molecular analogue of the ubiquitous photonic dual-rail encoding.

Our simplified error-correction conditions against prominent physical noise are often underconstrained, yielding other families of codes which correct against the same errors. {A natural extension of our work recently appeared in the form of approximate \AE\ codes~\cite{furey2024strategies}}. We anticipate that our noise-model simplification will yield a blueprint for code designs tailored to the specifics of a given atom or molecule.

\begin{acknowledgments}
  We thank
  Ivan H.\ Deutsch,
  Philipp Schindler,
  Aditya Sivakumar, and Thomas Dellaert
  for inspiring discussions.
  SPJ and VVA acknowledge financial support from NSF QLCI grant OMA-2120757.
  ERH and WCC acknowledge support from NSF (PHY-2110421, PHY-2207985 and OMA-2016245), AFOSR (130427-5114546 and FA9550-20-1-0323), and ARO (W911NF-20-1-0037 and W911NF-19-1-0297).   
  Contributions to this work by NIST, an agency of the US government, are not subject to US copyright.
  Any mention of commercial products does not indicate endorsement by NIST.
  VVA thanks Olga Albert and Ryhor Kandratsenia for providing daycare support throughout this work.
\end{acknowledgments}

\onecolumngrid
\appendix
\renewcommand{\tocname}{Appendices}
\pagebreak

\section{A. Non-locality of rigid body shift noise} 
\setcounter{section}{1} 

Consider the case of a planar rotor~\cite[Sec.~IV]{albert_robust_2020}, relevant to the case of a heteronuclear diatomic molecule confined to rotate in a plane.
The Hilbert space is described by the angular position states $\{\ket{\phi} : \phi \in [0,2\pi) \}$, or equivalently, the angular momentum  states $\{\ket{\ell} : \ell \in \mathbb{Z} \}$.

A natural decomposition of any trace-class operator $\hat{O}$ acting on the planar rotor space is
\begin{equation}\label{eqn:decompose-u1-operator}
    \hat{O} = \sum_{\ell\in\mathbb{Z}}\int_{0}^{2\pi} d\phi \,c_{\ell}(\phi)e^{i \ell \hat{\theta}} e^{-i \phi \hat{L}}\, \, .
\end{equation}
The position shift operators $e^{-i \phi \hat{L}} = \int_{0}^{2\pi}d\phi^\prime \, \ket{\phi+\phi^\prime}\bra{\phi^\prime}$ and the angular momentum shift operators $e^{i \ell \hat{\theta}} = \sum_{\ell^\prime \in \mathbb{Z}}\ket{\ell+\ell^\prime}\bra{\ell^\prime}$
form a complete basis of operators in this space.

Let us apply the above decomposition to a general angular momentum transition,    \begin{eqnarray}
        \hat{\sigma}_{mn} = \ket{m}\bra{n} &=& \sum_{\ell\in\mathbb{Z}}\int_{0}^{2\pi} d\phi \, c_{\ell}(\phi)e^{i \ell \hat{\theta}} e^{-i \phi \hat{L}}\, \,   \textrm{with} \\
        c_{\ell}(\phi) &=& \delta_{\ell,m-n} \dfrac{e^{i n \phi}}{2\pi}\, \, .
    \end{eqnarray}
    The weight of the $\phi$-position shift $|c_{m-n}(\phi)|$ is independent of $\phi$ i.e. there is uniform, non-zero contribution from \textit{all} angular position shift operators.
    Thus, a transition noise operator does not only contain small shifts in the angular position, but contains equal contributions from shifts of all possible amplitudes.
    In particular, the noise operators contains significant contributions from large angular position shifts and is hence, \textit{non-local} in the phase space of the rotor.

    Similar (but no longer orthonormal~\cite{albert_robust_2020}) decomposition into angular position shifts $(\hat{X}_S)$ and angular momentum shifts $(\hat{Y}^\ell_{m})$ is possible for the general diatomic molecule.
    We describe one such transition operator here:
    \begin{eqnarray}
        \ket{\ell,m-1}\bra{\ell,m}=\int_{ SO(3)}d\mu(S)\sum_{\substack{\ell^\prime\in\mathbb{Z}^+ \\ -\ell^\prime \leq m^\prime\leq \ell^\prime}}\mathcal{E}^{\ell^\prime}_{m^\prime}(S)\hat{Y}^{\ell^\prime}_{m^\prime}\hat{X}_S\, \,   \textrm{with} \\
        \mathcal{E}^{\ell^\prime}_{m^\prime n^\prime}(S) = \sqrt{\frac{2\ell^\prime+1}{4\pi}}D^{\ell}_{m-1,m-m^\prime}(S)C^{\ell,m\star}_{\ell^\prime,m^\prime;\ell,m-m^\prime}C^{ \ell,0\star}_{\ell^\prime,0;\ell,0}\, \, .
    \end{eqnarray}
    The actions of position shift operators $\hat{X}_S$ and the angular momentum shift operators $\hat{Y}^\ell_{m}$ are defined by their action on the position basis states $\ket{v}$
    \begin{eqnarray}
        \hat{X}_S\ket{v} &=& \ket{Sv}  \\
        \hat{Y}^\ell_{m}\ket{v} &=& {Y}^\ell_{m}(v)\ket{v}~,
    \end{eqnarray}
    where ${Y}^\ell_{m}(v)$ is the spherical harmonic evaluated at $v$,
    $C^{a,b}_{c,d;e,f}$ are the Clebsch-Gordan coefficients, and $\mu(S)$ is the Haar measure on $SO(3)$.
    The actual order-$r$ transition noise operators as described in the main text are
    \begin{equation}
        \hat{K}_{\delta m=-1}^{r(\delta \ell=0)} \propto \sum_{|m|\leq J}  C^{\ell,m-1}_{\ell,m;r,-1}\ket{\ell,m-1}\bra{\ell,m}\,.
    \end{equation}
    The relevant weights of the shift operators take the form
    \begin{equation}
    \mathcal{E}^{\ell^\prime}_{m^\prime }(S) \propto \sum_{|m|\leq \ell_0}\sqrt{\frac{2\ell^\prime+1}{4\pi}} C^{\ell,m-1}_{\ell,m;r,-1} C^{ \ell,m\star}_{\ell^\prime,m^\prime;\ell,m-m^\prime}C^{ \ell,0\star}_{\ell^\prime,0;\ell,0} D^{\ell}_{m-1,m-m^\prime}(S) \,.
    \end{equation}

    As an example, Fig. \ref{fig:appxE} shows the absolute value of the weights $|\mathcal{E}^2_{-1}(S)|$ for the order-$2$ transition operator acting on $J=2$ level of the system as a function of the angular shift about an axis perpendicular to the axis of the molecule.
    These weights reveal that the transition noise operators for the general diatomic molecule also contain significant contributions from large shifts in the angular orientation, causing molecular codes~\cite{albert_robust_2020} to fail against them.
    \begin{figure}[t!]
    \centering
    \includegraphics[width=0.5\columnwidth]{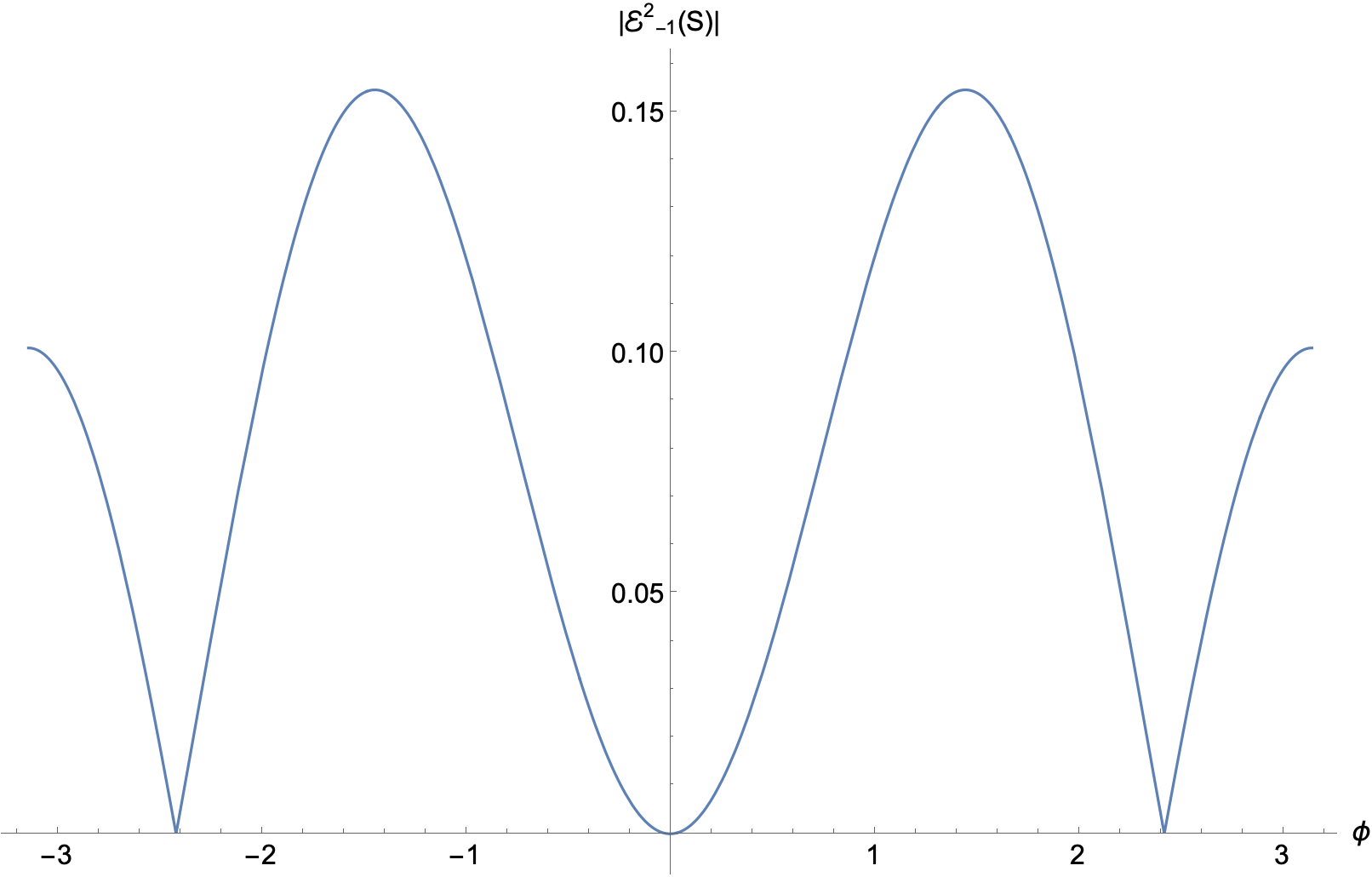}
    \caption{The weights $|\mathcal{E}^2_{-1}(S)|$ for the order-$2$ transition operator $\hat{K}_{-1}^{2( 0)}$ acting on $J=2$ level of the system as a function of the angle of rotation $\phi$ about the axis $(1,0,0)$ which is perpendicular to the symmetry axis of the molecule.
    The angle $\phi=0$ represents zero shift, and $\phi = \pm \pi$ is the maximum possible shift in the angular orientation of the rotor.
    }
    \label{fig:appxE}
\end{figure}

\section{B. No-go theorem}
\setcounter{section}{2}
We prove the following no-go theorem:
\emph{Given a 
Hilbert space with countable basis and orthogonal vectors $\ket{\psi}$ and $\ket{\phi}$, an error channel containing the Kraus operator $\hat{E} = \ket{\phi}\bra{\psi}$ is correctable only if the codespace is not supported by $\ket{\psi}$.}

\begin{proof}
    Proof is simple by checking the KL conditions for this error operator. Consider the two general encoded states
    \begin{equation}
    \ket{\bar{0}}={\textstyle \sum_{i}}\alpha_{i}\ket{\psi_i}\quad\text{and}\quad\ket{\bar{1}}={\textstyle \sum_{i}}\beta_{i}\ket{\psi_i},
\end{equation}
    with the sum running over the Hilbert space basis $\{\ket{\psi_i}\}$ and set $\ket{\psi_0}=\ket{\psi}$ without loss of generality. To be valid code states, the following conditions need to be satisfied simultaneously given \(\hat{E}^\dagger \hat{E} = |\psi_0\rangle\langle\psi_0|\). 
    \begin{align}
    |\alpha_{0}|^{2}=\bra{\bar{0}}\hat{E}^{\dagger}\hat{E}\ket{\bar{0}}&\overset{!}{=}\bra{\bar{1}}\hat{E}^{\dagger}\hat{E}\ket{\bar{1}}=|\beta_{0}|^{2}\label{eqn:nogocond1}\\\bra{\bar{0}}\hat{E}^{\dagger}\hat{E}\ket{\bar{1}}=\beta_{0}\alpha_{0}^{\star}&\overset{!}{=}0~,\label{eqn:nogocond2}    \end{align}
    where we denote satisfaction by the symbol ``\(\overset{!}{=}\)''.
  The only solution to Eqn.~\ref{eqn:nogocond1} and Eqn.~\ref{eqn:nogocond2} is $\alpha_0=\beta_0=0$ i.e. the code states having zero support over the state $\ket{\psi}$.
\end{proof}

\section{C. Protection against dephasing}
\setcounter{section}{3}

We concentrate on identifying code states designed to protect against dephasing errors up to order $k$. These code states are confined to a single $\ell$ manifold, and we assume a separation such that the off-diagonal KL conditions (i.e. $i \neq j$) are trivially fulfilled. The remaining KL constraints are 
\begin{equation}
\bra{\bar{0}}\hat{m}^r\ket{\bar{0}}-\bra{\bar{1}}\hat{m}^r\ket{\bar{1}}=0
\end{equation}
for all $r$ upto $2n$. Considering the general code states 
\begin{equation}
    \ket{\bar{0}}={\textstyle \sum_{m}}\alpha_{m}\ket{\ell,m}\quad\text{and}\quad\ket{\bar{1}}={\textstyle \sum_{m}}\beta_{m}\ket{\ell,m},
\end{equation}these constraints simply become linear equations
\begin{equation}
   \left\{ \sum_{m=-\ell}^{\ell}\left(|\alpha_m|^2 - |\beta_m|^2\right) m^r = 0 \right\}_{0\leq r \leq 2n}.
\end{equation}
This is a system of $2n$ linear equations in the $4\ell + 2$ variables $|\alpha_i|^2, |\beta_j|^2$. One might initially consider the solution $\alpha_i = \beta_i \forall , i$, but quickly realizes that it fails to meet the necessary requirements for our ansatz and is not valid. Due to the constraints on the ansatz, the actual number of variables involved is generally much smaller than $4\ell + 2$. To illustrate, we explicitly outline the system of equations for spontaneous decay arising from the symmetric ansatz.
\begin{equation}
    \ket{\bar{0}}=\alpha_{-m_1}\ket{\ell,-m_1}+\alpha_{m_1}\ket{\ell,m_1}\quad\text{and}\quad\ket{\bar{1}}=\beta_{-m_2}\ket{\ell,-m_2}+\beta_{0}\ket{\ell,0}+\beta_{m_2}\ket{\ell,m_2}.
\end{equation}
Here, instead of $4\ell + 2$ variables, there are only $5$, independent of $\ell$ (provided $\ell\geq 6$). The constraints to be satisfied are
\begin{eqnarray}\label{eqn:linear-system-spontaneous-decay-symmetric}
    \left\{|\alpha_{-m_1}|^2(-m_1)^i +|\alpha_{m_1}|^2 m_1^i- |\beta_{0}|^2 \delta_{0,i} - |\beta_{-m_2}|^2 (-m_2)^i - |\beta_{m_2}|^2 m_2^i = 0 \right\}_{0\leq i\leq 2} .
\end{eqnarray}

This is a system of \emph{three} linear equations in \emph{five} variables admitting infinitely many solutions, one of which is presented as the symmetric $n=1$ code in the main text.

Our method enables the construction of such error-correcting code families for protection against errors of any order. We illustrate with a countersymmetric code family, designed to correct upto order 2 errors.
\begin{eqnarray}
    \ket{\bar{0}}&=&{\textstyle\sqrt{\frac{m_1 \left(m_1-m_2\right) m_2}{\left(m_1-m_3\right) \left(m_1-m_2+m_3\right) \left(m_2+m_3\right)}}}\ket{J,-m_3}+{\textstyle\sqrt{\frac{m_2 \left(m_2-m_3\right) m_3}{\left(m_1+m_2\right) \left(m_1-m_3\right) \left(m_1-m_2+m_3\right)}}}\ket{J,-m_1}\nonumber\\&&+{\textstyle\sqrt{\frac{m_1 m_3 \left(m_1+m_3\right)}{\left(m_1+m_2\right) \left(m_1-m_2+m_3\right) \left(m_2+m_3\right)}}}\ket{J,m_2} \nonumber\\
    \ket{\bar{1}}&=&{\textstyle\sqrt{\frac{m_1 m_3 \left(m_1+m_3\right)}{\left(m_1+m_2\right) \left(m_1-m_2+m_3\right) \left(m_2+m_3\right)}}}\ket{J,-m_2}+{\textstyle\sqrt{\frac{m_2 \left(m_2-m_3\right) m_3}{\left(m_1+m_2\right) \left(m_1-m_3\right) \left(m_1-m_2+m_3\right)}}}\ket{J,m_1}\nonumber \\&&+{\textstyle\sqrt{\frac{m_1 \left(m_1-m_2\right) m_2}{\left(m_1-m_3\right) \left(m_1-m_2+m_3\right) \left(m_2+m_3\right)}}}\ket{J,m_3}
\end{eqnarray}
where $m_3-m_1,m_2-m_1\geq 5$ and $m_1\geq 2$.
Lastly, our approach is not limited to systems with inherent symmetries; it can be applied to construct codes using any ansatz that meets the separation requirements. As a comprehensive illustration, we showcase the general compact code capable of correcting up to order $1$ errors.
\begin{align}
\ket{\bar{0}}&={\textstyle\sqrt{\frac{\left(m_3-m_2\right) \left(m_4-m_2\right)}{\left(m_2-m_1\right) \left(m_1+m_2-m_3-m_4\right)}}}\ket{\ell,m_1}+{\textstyle\sqrt{\frac{\left(m_1-m_3\right) \left(m_1-m_4\right)}{\left(m_1-m_2\right) \left(m_1+m_2-m_3-m_4\right)}}}\ket{\ell,m_2}\nonumber\\
\ket{\bar{1}}&={\textstyle\sqrt{\frac{\left(m_1-m_4\right) \left(m_2-m_4\right)}{\left(m_1+m_2-m_3-m_4\right) \left(m_3-m_4\right)}}}\ket{\ell,m_3}+{\textstyle\sqrt{\frac{\left(m_1-m_3\right) \left(m_3-m_2\right)}{\left(m_1+m_2-m_3-m_4\right) \left(m_3-m_4\right)}}}\ket{\ell,m_4}.
\end{align}
where the separation requirement are met if $|m_i-m_j|\geq 3\, \forall\, i\neq j$. This family admits the counter-symmetric $n=1$ code as a member upon setting $m_1=-m_1, m_3=-m_2$ and $m_4=m_1$.

\section{D. Theorems concerning $\hat{Y}$ matrix elements}
\setcounter{section}{4}
Using an ansatz with sufficient separation between supported states, the only KL conditions for higher order transition errors which are not trivially satisfied include terms of the form $\bra{J_0,m}\hat{K}^{r_1\dagger}\hat{K}^{r_2}\ket{J_0,m}$ (see main text for definition of $\hat{K}^r$) which are proportional to products of $\hat{Y}$ operators $\bra{\ell_0+\delta \ell,m+\delta m}\hat{Y}^{r_1}_{\delta m}\ket{\ell_0,m}
\bra{\ell_0+\delta \ell,m+\delta m}\hat{Y}^{r_2}_{\delta m}\ket{\ell_0,m}$. Then, the following claim is adequate to prove that such KL conditions reduce to the KL conditions for the generalized dephasing operator set $\left\{\hat{m}^k\, | \, 0\leq k\leq n\right\}$ for any $n\geq \max(r_1,r_2)$:
\begin{claim} \label{claim:yhat-product}
$\bra{\ell_0+\delta \ell,m+\delta m}\hat{Y}^{r_1}_{\delta m}\ket{\ell_0,m}
\bra{\ell_0+\delta \ell,m+\delta m}\hat{Y}^{r_2}_{\delta m}\ket{\ell_0,m} = poly(r_1 + r_2, m)$ where \(poly(r,m)\) denotes a general polynomial of degree $r$ in variable $m$ and $\hat{Y}$ matrix elements are as defined in \cite[TABLE V]{albert_robust_2020}.
\end{claim}
We note some other results required to prove the stated claim. The matrix elements of $\hat{Y}$ can be written as a product of Clebsch-Gordan(CG) coefficients
\begin{equation}
    \bra{\ell,m+\delta m}\hat{Y}^r_{\delta m}\ket{\ell_0,m} = {\textstyle\sqrt{\frac{(2r+1)(2\ell_0+1)}{4\pi(2\ell+1)}}}C^{\ell,m+\delta m}_{r,\delta m; \ell_0,m}C^{\ell,0}_{r,0; \ell_0,0} \propto C^{\ell,m+\delta m}_{\ell_0,m; r,\delta m}
\end{equation}
where the $m$ dependence lies entirely in one CG coefficient. 
\begin{observation} \label{obs:extremeCGaresqrtpolynomials}
    For integer or half integer $\ell_0$, all Clebsch-Gordan coefficients of the form $C_{\ell_{0},m;r,\pm r}^{\ell,m\pm r}$ and $C_{\ell_{0},m;r,\delta m}^{\ell\pm r,m+\delta m}$ are square roots of real polynomials of degree $2r$ in $m$.
\end{observation}
This observation is verified simply by looking at the expressions for the mentioned CG coefficients \cite[Chapter 8]{varshalovich_quantum_1988}. 
\theorem{For $r\geq |\delta m|$, $C^{\ell,m+\delta m}_{r,\delta m; \ell_0,m}=poly(m,r-|\delta m|)C^{\ell,m+\delta m}_{|\delta m|,\delta m; \ell_0,m}$} .\label{thm:recursive-cg}

\begin{proof}
  Note the identity involving Clebsch-Gordan coefficients (8.6.5 \cite{varshalovich_quantum_1988})
    \begin{eqnarray} \label{eqn:versalovich-CGidentity}
        C_{J_{0},m;r,\delta m}^{J,m+\delta m}&=&\left[2m+\textstyle{\frac{\delta m}{r(r+1)}}\{r(r+1)+J_{0}(J_{0}+1)-J(J+1)\}\right]^{-1}\nonumber \\
        &&\biggl[{\textstyle\frac{1}{r(2r+1)}}\left\{\left(r^{2}-\delta m^{2}\right)(-r+J_{0}+J+1)  (r-J_{0}+J)(r+J_{0}-J)(r+J_{0}+J+1)\right\}^{\frac{1}{2}}C_{J_{0},m;r-1,\delta m}^{J,m+\delta m} +\nonumber \\
        &&{\textstyle\frac{1}{(r+1)(2r+1)}}\left\{\left[(r+1)^{2}-\delta m^{2}\right](-r+J_{0}+J)(r-J_{0}+J+1)(r+J_{0}-J+1)(r+J_{0}+J+2)\right\} ^{\frac{1}{2}} \nonumber \\&& C_{J_{0},m;r+1,\delta m}^{J,m+\delta m}\biggr].
    \end{eqnarray}
 For the sake of clarity, we substitute $C_{J_{0},m;r,\delta m}^{J,m+\delta m}$ with $C_r(m)$ to label the rank $r$ symbol as a function of $m$. Then, 
 \begin{eqnarray}\label{eqn:recursion-CGrankr}
    C_{r}(m) = poly_1(m,1)C_{r-1}(m) + poly_2(m,0)C_{r-2}.
 \end{eqnarray}
 We also observe that for every $poly_1,poly_2$, the product of the two is some $poly_3$ such that $poly_1(m,a)poly_2(m,b)=poly_3(m,a+b)$.
Using these relations, we inductively prove our theorem. The statement to prove is that for every $r\geq |\delta m|$, $C_r(m)=poly(m,r-|\delta m|)C_{|\delta m|}(m)$ .

\textbf{Base case:} $C_{|\delta m|} = C_{|\delta m|}(m)=poly(m,0)C_{|\delta m|}$ and $C_{|\delta m| +1}(m) = poly(m,1)C_{|\delta m|}(m)$ using Eq. \eqref{eqn:recursion-CGrankr}.

\textbf{Inductive step:} Assume that $C_{r-1}(m) = poly_1(m,r-1-|\delta m|)C_{|\delta m|}(m)$ and $C_{r-2}(m) = poly_2(m,r-2-|\delta m|)C_{|\delta m|}(m)$, then
\begin{eqnarray}
    C_r(m) &=& \tilde{poly_1}(m,1)C_{r-1}(m) + \tilde{poly_2}(m,0)C_{r-2}(m)    \nonumber \\
           &=& \left(\tilde{poly_1}(m,1)poly_1(m,r-1-|\delta m|)+ \tilde{poly_2}(m,0)poly_2(m,r-2-|\delta m|)\right)C_{|\delta m|}(m) \nonumber \\
           &=& poly(m,r-|\delta m|)C_{|\delta m|}(m).
\end{eqnarray}
Combining the base case and the inductive step, the statement is proved to hold for $r\geq |\delta m|$.
\end{proof}

\theorem{
    All Clebsch-Gordan coefficients $C^{\ell,m+\delta m}_{r,\delta m; \ell_0,m}$ can be formulated as square roots of real polynomials of degree $2r$ in $m$.
}
\begin{proof}
    For clarity, we will denote by $C_r^{\delta \ell, \delta m}(m)$ the Clebsch-Gordan coefficient $C^{\ell_0+\delta \ell,m+\delta m}_{\ell_0,m;r,\delta m}$ as a function of $m$.
    The theorem can be proved by induction. The statement to prove for all CG coefficients is $C_r^{\delta \ell,\delta m}(m)=\sqrt{poly(m,2r)}$.
    
    \textbf{Base case:} $C_0^{\delta \ell=0,\delta m=0}(m)= \sqrt{poly(m,0)}$ by observation \ref{obs:extremeCGaresqrtpolynomials}.

    \textbf{Inductive step:} Assume that $C_k^{\delta \ell, \delta m}(m)= \sqrt{poly(m,2k)}\, \forall \, |\delta \ell|, |\delta m| \leq k \, \forall \, k$ upto $r-1$. 
    First, consider the coefficients $C^{\delta \ell=\pm r,\delta m}_r(m)$ and $C^{\delta \ell,\delta m = \pm r}_r(m)$. These satisfy the statement by use of Observation \ref{obs:extremeCGaresqrtpolynomials}. Next, consider all other coefficients $C^{\delta \ell, \delta m}_r(m)$ such that $\delta \ell \neq \pm r, \delta m \neq \pm r$. Then using Theorem \ref{thm:recursive-cg} and Observation \ref{obs:extremeCGaresqrtpolynomials}, these satisfy 
\begin{eqnarray}
    C^{\delta \ell, \delta m}_r(m) &=& \sqrt{poly(m,2r-2|\delta m|)}C^{\delta \ell,\delta m}_{|\delta m|} \nonumber \\
    &=& \sqrt{poly(m,2r)}.
\end{eqnarray}
Combining the base case and the inductive step, the statement is proved to hold for all CG coefficients.
\end{proof}
\normalfont{We are finally ready to prove claim \ref{claim:yhat-product}}.
\begin{proof}
    With the use the stated theorems, the product of the $\hat{Y}$ matrix elements in claim \ref{claim:yhat-product} simplifies as 
\begin{eqnarray}
\bra{\ell_0+\delta \ell,m+\delta m}\hat{Y}^{r_1}_{\delta m}\ket{\ell_0,m}
\bra{\ell_0+\delta \ell,m+\delta m}\hat{Y}^{r_2}_{\delta m}\ket{\ell_0,m} &=& poly_1(m,r_1-|\delta m|) \nonumber\\
     &\times& poly_2(m,r_2-|\delta m|) \nonumber \\
     &\times& \left(\sqrt{poly_3(m,2\delta m)}\right)^2 \nonumber \\
     &=& poly(m, r_1 + r_2 ).
\end{eqnarray}
\end{proof}

\section{E. Binomial like \AE\ codes}
\setcounter{section}{5}
Here, we show that the binomial-like qudit codestates, as formulated in the main text protect against dephasing errors upto order $n$. It is sufficient to show $\bra{\bar{0}}\hat{m}^r\ket{\bar{0}}-\bra{\bar{1}}\hat{m}^r\ket{\bar{1}}=0 \, \forall \, 0\leq r \leq 2n$:
\begin{eqnarray}
    \bra{\bar{0}}\hat{m}^k\ket{\bar{0}}-\bra{\bar{1}}\hat{m}^k\ket{\bar{1}} &=& \sum_{\text{even }k=0}^{2n+1}\binom{2n+1}{k}\left[-m_{0}+k(2n+1)\right]^{r}-\sum_{\text{odd }k=0}^{2n+1}\binom{2n+1}{k}\left[-m_{0}+k(2n+1)\right]^{r} \nonumber \\
    &=& \sum_{k=0}^{2n+1}\binom{2n+1}{k}(-1)^k\left[-m_{0}+k(2n+1)\right]^{r} \nonumber \\
    &=& \sum_{l=0}^{r}\binom{r}{l}(-m_0)^{r-l}(2n+1)^l \left\{\sum_{k=0}^{2n+1}\binom{2n+1}{k}(-1)^kk^l\right\} \nonumber \\
\end{eqnarray}
where it is a well known binomial identity that the sum in the curly braces is zero $ \forall \, 0\leq l\leq 2n$ i.e. $\forall \, 0\leq r\leq 2n$, the entire sum vanishes.

\vspace{1in}

\twocolumngrid
\bibliography{references}

\begin{thebibliography}{66}%
\makeatletter
\providecommand \@ifxundefined [1]{%
 \@ifx{#1\undefined}
}%
\providecommand \@ifnum [1]{%
 \ifnum #1\expandafter \@firstoftwo
 \else \expandafter \@secondoftwo
 \fi
}%
\providecommand \@ifx [1]{%
 \ifx #1\expandafter \@firstoftwo
 \else \expandafter \@secondoftwo
 \fi
}%
\providecommand \natexlab [1]{#1}%
\providecommand \enquote  [1]{``#1''}%
\providecommand \bibnamefont  [1]{#1}%
\providecommand \bibfnamefont [1]{#1}%
\providecommand \citenamefont [1]{#1}%
\providecommand \href@noop [0]{\@secondoftwo}%
\providecommand \href [0]{\begingroup \@sanitize@url \@href}%
\providecommand \@href[1]{\@@startlink{#1}\@@href}%
\providecommand \@@href[1]{\endgroup#1\@@endlink}%
\providecommand \@sanitize@url [0]{\catcode `\\12\catcode `\$12\catcode
  `\&12\catcode `\#12\catcode `\^12\catcode `\_12\catcode `\%12\relax}%
\providecommand \@@startlink[1]{}%
\providecommand \@@endlink[0]{}%
\providecommand \url  [0]{\begingroup\@sanitize@url \@url }%
\providecommand \@url [1]{\endgroup\@href {#1}{\urlprefix }}%
\providecommand \urlprefix  [0]{URL }%
\providecommand \Eprint [0]{\href }%
\providecommand \doibase [0]{https://doi.org/}%
\providecommand \selectlanguage [0]{\@gobble}%
\providecommand \bibinfo  [0]{\@secondoftwo}%
\providecommand \bibfield  [0]{\@secondoftwo}%
\providecommand \translation [1]{[#1]}%
\providecommand \BibitemOpen [0]{}%
\providecommand \bibitemStop [0]{}%
\providecommand \bibitemNoStop [0]{.\EOS\space}%
\providecommand \EOS [0]{\spacefactor3000\relax}%
\providecommand \BibitemShut  [1]{\csname bibitem#1\endcsname}%
\let\auto@bib@innerbib\@empty
\bibitem [{\citenamefont {Yan}\ \emph {et~al.}(2013)\citenamefont {Yan},
  \citenamefont {Moses}, \citenamefont {Gadway}, \citenamefont {Covey},
  \citenamefont {Hazzard}, \citenamefont {Rey}, \citenamefont {Jin},\ and\
  \citenamefont {Ye}}]{yan_observation_2013}%
  \BibitemOpen
  \bibfield  {author} {\bibinfo {author} {\bibfnamefont {B.}~\bibnamefont
  {Yan}}, \bibinfo {author} {\bibfnamefont {S.~A.}\ \bibnamefont {Moses}},
  \bibinfo {author} {\bibfnamefont {B.}~\bibnamefont {Gadway}}, \bibinfo
  {author} {\bibfnamefont {J.~P.}\ \bibnamefont {Covey}}, \bibinfo {author}
  {\bibfnamefont {K.~R.~A.}\ \bibnamefont {Hazzard}}, \bibinfo {author}
  {\bibfnamefont {A.~M.}\ \bibnamefont {Rey}}, \bibinfo {author} {\bibfnamefont
  {D.~S.}\ \bibnamefont {Jin}},\ and\ \bibinfo {author} {\bibfnamefont
  {J.}~\bibnamefont {Ye}},\ }\bibfield  {title} {\bibinfo {title} {Observation
  of dipolar spin-exchange interactions with lattice-confined polar
  molecules},\ }\href {https://doi.org/10.1038/nature12483} {\bibfield
  {journal} {\bibinfo  {journal} {Nature}\ }\textbf {\bibinfo {volume} {501}},\
  \bibinfo {pages} {521} (\bibinfo {year} {2013})},\ \bibinfo {note} {number:
  7468 Publisher: Nature Publishing Group}\BibitemShut {NoStop}%
\bibitem [{\citenamefont {Ni}\ \emph {et~al.}(2018)\citenamefont {Ni},
  \citenamefont {Rosenband},\ and\ \citenamefont {Grimes}}]{ni_dipolar_2018}%
  \BibitemOpen
  \bibfield  {author} {\bibinfo {author} {\bibfnamefont {K.-K.}\ \bibnamefont
  {Ni}}, \bibinfo {author} {\bibfnamefont {T.}~\bibnamefont {Rosenband}},\ and\
  \bibinfo {author} {\bibfnamefont {D.~D.}\ \bibnamefont {Grimes}},\ }\bibfield
   {title} {\bibinfo {title} {Dipolar exchange quantum logic gate with polar
  molecules},\ }\href {https://doi.org/10.1039/C8SC02355G} {\bibfield
  {journal} {\bibinfo  {journal} {Chem. Sci.}\ }\textbf {\bibinfo {volume}
  {9}},\ \bibinfo {pages} {6830} (\bibinfo {year} {2018})}\BibitemShut
  {NoStop}%
\bibitem [{\citenamefont {Seesselberg}\ \emph {et~al.}(2018)\citenamefont
  {Seesselberg}, \citenamefont {Luo}, \citenamefont {Li}, \citenamefont
  {Bause}, \citenamefont {Kotochigova}, \citenamefont {Bloch},\ and\
  \citenamefont {Gohle}}]{seeselberg_extending_2018}%
  \BibitemOpen
  \bibfield  {author} {\bibinfo {author} {\bibfnamefont {F.}~\bibnamefont
  {Seesselberg}}, \bibinfo {author} {\bibfnamefont {X.-Y.}\ \bibnamefont
  {Luo}}, \bibinfo {author} {\bibfnamefont {M.}~\bibnamefont {Li}}, \bibinfo
  {author} {\bibfnamefont {R.}~\bibnamefont {Bause}}, \bibinfo {author}
  {\bibfnamefont {S.}~\bibnamefont {Kotochigova}}, \bibinfo {author}
  {\bibfnamefont {I.}~\bibnamefont {Bloch}},\ and\ \bibinfo {author}
  {\bibfnamefont {C.}~\bibnamefont {Gohle}},\ }\bibfield  {title} {\bibinfo
  {title} {Extending {Rotational} {Coherence} of {Interacting} {Polar}
  {Molecules} in a {Spin}-{Decoupled} {Magic} {Trap}},\ }\href
  {https://doi.org/10.1103/PhysRevLett.121.253401} {\bibfield  {journal}
  {\bibinfo  {journal} {Phys. Rev. Lett.}\ }\textbf {\bibinfo {volume} {121}},\
  \bibinfo {pages} {253401} (\bibinfo {year} {2018})}\BibitemShut {NoStop}%
\bibitem [{\citenamefont {Burchesky}\ \emph {et~al.}(2021)\citenamefont
  {Burchesky}, \citenamefont {Anderegg}, \citenamefont {Bao}, \citenamefont
  {Yu}, \citenamefont {Chae}, \citenamefont {Ketterle}, \citenamefont {Ni},\
  and\ \citenamefont {Doyle}}]{burchesky_rotational_2021}%
  \BibitemOpen
  \bibfield  {author} {\bibinfo {author} {\bibfnamefont {S.}~\bibnamefont
  {Burchesky}}, \bibinfo {author} {\bibfnamefont {L.}~\bibnamefont {Anderegg}},
  \bibinfo {author} {\bibfnamefont {Y.}~\bibnamefont {Bao}}, \bibinfo {author}
  {\bibfnamefont {S.~S.}\ \bibnamefont {Yu}}, \bibinfo {author} {\bibfnamefont
  {E.}~\bibnamefont {Chae}}, \bibinfo {author} {\bibfnamefont {W.}~\bibnamefont
  {Ketterle}}, \bibinfo {author} {\bibfnamefont {K.-K.}\ \bibnamefont {Ni}},\
  and\ \bibinfo {author} {\bibfnamefont {J.~M.}\ \bibnamefont {Doyle}},\
  }\bibfield  {title} {\bibinfo {title} {Rotational {Coherence} {Times} of
  {Polar} {Molecules} in {Optical} {Tweezers}},\ }\href
  {https://doi.org/10.1103/PhysRevLett.127.123202} {\bibfield  {journal}
  {\bibinfo  {journal} {Physical Review Letters}\ }\textbf {\bibinfo {volume}
  {127}},\ \bibinfo {pages} {123202} (\bibinfo {year} {2021})},\ \bibinfo
  {note} {publisher: American Physical Society}\BibitemShut {NoStop}%
\bibitem [{\citenamefont {Cairncross}\ \emph {et~al.}(2021)\citenamefont
  {Cairncross}, \citenamefont {Zhang}, \citenamefont {Picard}, \citenamefont
  {Yu}, \citenamefont {Wang},\ and\ \citenamefont
  {Ni}}]{cairncross_assembly_2021}%
  \BibitemOpen
  \bibfield  {author} {\bibinfo {author} {\bibfnamefont {W.~B.}\ \bibnamefont
  {Cairncross}}, \bibinfo {author} {\bibfnamefont {J.~T.}\ \bibnamefont
  {Zhang}}, \bibinfo {author} {\bibfnamefont {L.~R.}\ \bibnamefont {Picard}},
  \bibinfo {author} {\bibfnamefont {Y.}~\bibnamefont {Yu}}, \bibinfo {author}
  {\bibfnamefont {K.}~\bibnamefont {Wang}},\ and\ \bibinfo {author}
  {\bibfnamefont {K.-K.}\ \bibnamefont {Ni}},\ }\bibfield  {title} {\bibinfo
  {title} {Assembly of a {Rovibrational} {Ground} {State} {Molecule} in an
  {Optical} {Tweezer}},\ }\href
  {https://doi.org/10.1103/PhysRevLett.126.123402} {\bibfield  {journal}
  {\bibinfo  {journal} {Physical Review Letters}\ }\textbf {\bibinfo {volume}
  {126}},\ \bibinfo {pages} {123402} (\bibinfo {year} {2021})},\ \bibinfo
  {note} {publisher: American Physical Society}\BibitemShut {NoStop}%
\bibitem [{\citenamefont {Gregory}\ \emph {et~al.}(2021)\citenamefont
  {Gregory}, \citenamefont {Blackmore}, \citenamefont {Bromley}, \citenamefont
  {Hutson},\ and\ \citenamefont {Cornish}}]{gregory_robust_2021}%
  \BibitemOpen
  \bibfield  {author} {\bibinfo {author} {\bibfnamefont {P.~D.}\ \bibnamefont
  {Gregory}}, \bibinfo {author} {\bibfnamefont {J.~A.}\ \bibnamefont
  {Blackmore}}, \bibinfo {author} {\bibfnamefont {S.~L.}\ \bibnamefont
  {Bromley}}, \bibinfo {author} {\bibfnamefont {J.~M.}\ \bibnamefont
  {Hutson}},\ and\ \bibinfo {author} {\bibfnamefont {S.~L.}\ \bibnamefont
  {Cornish}},\ }\bibfield  {title} {\bibinfo {title} {Robust storage qubits in
  ultracold polar molecules},\ }\href
  {https://doi.org/10.1038/s41567-021-01328-7} {\bibfield  {journal} {\bibinfo
  {journal} {Nature Physics}\ }\textbf {\bibinfo {volume} {17}},\ \bibinfo
  {pages} {1149} (\bibinfo {year} {2021})},\ \bibinfo {note} {number: 10
  Publisher: Nature Publishing Group}\BibitemShut {NoStop}%
\bibitem [{\citenamefont {Holland}\ \emph {et~al.}(2022)\citenamefont
  {Holland}, \citenamefont {Lu},\ and\ \citenamefont
  {Cheuk}}]{holland_-demand_2022}%
  \BibitemOpen
  \bibfield  {author} {\bibinfo {author} {\bibfnamefont {C.~M.}\ \bibnamefont
  {Holland}}, \bibinfo {author} {\bibfnamefont {Y.}~\bibnamefont {Lu}},\ and\
  \bibinfo {author} {\bibfnamefont {L.~W.}\ \bibnamefont {Cheuk}},\ }\href
  {https://doi.org/10.48550/arXiv.2210.06309} {\bibinfo {title} {On-{Demand}
  {Entanglement} of {Molecules} in a {Reconfigurable} {Optical} {Tweezer}
  {Array}}} (\bibinfo {year} {2022}),\ \bibinfo {note} {arXiv:2210.06309
  [cond-mat, physics:physics, physics:quant-ph]}\BibitemShut {NoStop}%
\bibitem [{\citenamefont {Christakis}\ \emph {et~al.}(2023)\citenamefont
  {Christakis}, \citenamefont {Rosenberg}, \citenamefont {Raj}, \citenamefont
  {Chi}, \citenamefont {Morningstar}, \citenamefont {Huse}, \citenamefont
  {Yan},\ and\ \citenamefont {Bakr}}]{christakis_probing_2023}%
  \BibitemOpen
  \bibfield  {author} {\bibinfo {author} {\bibfnamefont {L.}~\bibnamefont
  {Christakis}}, \bibinfo {author} {\bibfnamefont {J.~S.}\ \bibnamefont
  {Rosenberg}}, \bibinfo {author} {\bibfnamefont {R.}~\bibnamefont {Raj}},
  \bibinfo {author} {\bibfnamefont {S.}~\bibnamefont {Chi}}, \bibinfo {author}
  {\bibfnamefont {A.}~\bibnamefont {Morningstar}}, \bibinfo {author}
  {\bibfnamefont {D.~A.}\ \bibnamefont {Huse}}, \bibinfo {author}
  {\bibfnamefont {Z.~Z.}\ \bibnamefont {Yan}},\ and\ \bibinfo {author}
  {\bibfnamefont {W.~S.}\ \bibnamefont {Bakr}},\ }\bibfield  {title} {\bibinfo
  {title} {Probing site-resolved correlations in a spin system of ultracold
  molecules},\ }\href {https://doi.org/10.1038/s41586-022-05558-4} {\bibfield
  {journal} {\bibinfo  {journal} {Nature}\ }\textbf {\bibinfo {volume} {614}},\
  \bibinfo {pages} {64} (\bibinfo {year} {2023})},\ \bibinfo {note} {number:
  7946 Publisher: Nature Publishing Group}\BibitemShut {NoStop}%
\bibitem [{\citenamefont {Ertmer}\ \emph {et~al.}(1985)\citenamefont {Ertmer},
  \citenamefont {Blatt}, \citenamefont {Hall},\ and\ \citenamefont
  {Zhu}}]{ertmer_laser_1985}%
  \BibitemOpen
  \bibfield  {author} {\bibinfo {author} {\bibfnamefont {W.}~\bibnamefont
  {Ertmer}}, \bibinfo {author} {\bibfnamefont {R.}~\bibnamefont {Blatt}},
  \bibinfo {author} {\bibfnamefont {J.~L.}\ \bibnamefont {Hall}},\ and\
  \bibinfo {author} {\bibfnamefont {M.}~\bibnamefont {Zhu}},\ }\bibfield
  {title} {\bibinfo {title} {Laser {Manipulation} of {Atomic} {Beam}
  {Velocities}: {Demonstration} of {Stopped} {Atoms} and {Velocity}
  {Reversal}},\ }\href {https://doi.org/10.1103/PhysRevLett.54.996} {\bibfield
  {journal} {\bibinfo  {journal} {Physical Review Letters}\ }\textbf {\bibinfo
  {volume} {54}},\ \bibinfo {pages} {996} (\bibinfo {year} {1985})},\ \bibinfo
  {note} {publisher: American Physical Society}\BibitemShut {NoStop}%
\bibitem [{\citenamefont {Weinstein}\ \emph {et~al.}(1998)\citenamefont
  {Weinstein}, \citenamefont {deCarvalho}, \citenamefont {Guillet},
  \citenamefont {Friedrich},\ and\ \citenamefont
  {Doyle}}]{weinstein_magnetic_1998}%
  \BibitemOpen
  \bibfield  {author} {\bibinfo {author} {\bibfnamefont {J.~D.}\ \bibnamefont
  {Weinstein}}, \bibinfo {author} {\bibfnamefont {R.}~\bibnamefont
  {deCarvalho}}, \bibinfo {author} {\bibfnamefont {T.}~\bibnamefont {Guillet}},
  \bibinfo {author} {\bibfnamefont {B.}~\bibnamefont {Friedrich}},\ and\
  \bibinfo {author} {\bibfnamefont {J.~M.}\ \bibnamefont {Doyle}},\ }\bibfield
  {title} {\bibinfo {title} {Magnetic trapping of calcium monohydride molecules
  at millikelvin temperatures},\ }\href {https://doi.org/10.1038/25949}
  {\bibfield  {journal} {\bibinfo  {journal} {Nature}\ }\textbf {\bibinfo
  {volume} {395}},\ \bibinfo {pages} {148} (\bibinfo {year} {1998})},\ \bibinfo
  {note} {number: 6698 Publisher: Nature Publishing Group}\BibitemShut
  {NoStop}%
\bibitem [{\citenamefont {Mengel}\ and\ \citenamefont
  {Lucia}(2000)}]{mengel_helium_2000}%
  \BibitemOpen
  \bibfield  {author} {\bibinfo {author} {\bibfnamefont {M.}~\bibnamefont
  {Mengel}}\ and\ \bibinfo {author} {\bibfnamefont {F.~C.~D.}\ \bibnamefont
  {Lucia}},\ }\bibfield  {title} {\bibinfo {title} {Helium and {Hydrogen}
  {Induced} {Rotational} {Relaxation} of {H2CO} {Observed} at {Temperatures} of
  the {Interstellar} {Medium}},\ }\href {https://doi.org/10.1086/317108}
  {\bibfield  {journal} {\bibinfo  {journal} {The Astrophysical Journal}\
  }\textbf {\bibinfo {volume} {543}},\ \bibinfo {pages} {271} (\bibinfo {year}
  {2000})},\ \bibinfo {note} {publisher: IOP Publishing}\BibitemShut {NoStop}%
\bibitem [{\citenamefont {Campbell}\ \emph {et~al.}(2007)\citenamefont
  {Campbell}, \citenamefont {Tsikata}, \citenamefont {Lu}, \citenamefont {van
  Buuren},\ and\ \citenamefont {Doyle}}]{campbell_magnetic_2007}%
  \BibitemOpen
  \bibfield  {author} {\bibinfo {author} {\bibfnamefont {W.~C.}\ \bibnamefont
  {Campbell}}, \bibinfo {author} {\bibfnamefont {E.}~\bibnamefont {Tsikata}},
  \bibinfo {author} {\bibfnamefont {H.-I.}\ \bibnamefont {Lu}}, \bibinfo
  {author} {\bibfnamefont {L.~D.}\ \bibnamefont {van Buuren}},\ and\ \bibinfo
  {author} {\bibfnamefont {J.~M.}\ \bibnamefont {Doyle}},\ }\bibfield  {title}
  {\bibinfo {title} {Magnetic {Trapping} and {Zeeman} {Relaxation} of {NH}
  $({X}^3 {\Sigma}^-)$},\ }\href
  {https://doi.org/10.1103/PhysRevLett.98.213001} {\bibfield  {journal}
  {\bibinfo  {journal} {Physical Review Letters}\ }\textbf {\bibinfo {volume}
  {98}},\ \bibinfo {pages} {213001} (\bibinfo {year} {2007})},\ \bibinfo {note}
  {publisher: American Physical Society}\BibitemShut {NoStop}%
\bibitem [{\citenamefont {Steinecker}\ \emph {et~al.}(2016)\citenamefont
  {Steinecker}, \citenamefont {McCarron}, \citenamefont {Zhu},\ and\
  \citenamefont {DeMille}}]{steinecker_improved_2016}%
  \BibitemOpen
  \bibfield  {author} {\bibinfo {author} {\bibfnamefont {M.~H.}\ \bibnamefont
  {Steinecker}}, \bibinfo {author} {\bibfnamefont {D.~J.}\ \bibnamefont
  {McCarron}}, \bibinfo {author} {\bibfnamefont {Y.}~\bibnamefont {Zhu}},\ and\
  \bibinfo {author} {\bibfnamefont {D.}~\bibnamefont {DeMille}},\ }\bibfield
  {title} {\bibinfo {title} {Improved {Radio}-{Frequency} {Magneto}-{Optical}
  {Trap} of {SrF} {Molecules}},\ }\href
  {https://doi.org/10.1002/cphc.201600967} {\bibfield  {journal} {\bibinfo
  {journal} {ChemPhysChem}\ }\textbf {\bibinfo {volume} {17}},\ \bibinfo
  {pages} {3664} (\bibinfo {year} {2016})}\BibitemShut {NoStop}%
\bibitem [{\citenamefont {Kozyryev}\ \emph {et~al.}(2017)\citenamefont
  {Kozyryev}, \citenamefont {Baum}, \citenamefont {Matsuda}, \citenamefont
  {Augenbraun}, \citenamefont {Anderegg}, \citenamefont {Sedlack},\ and\
  \citenamefont {Doyle}}]{kozyryev_sisyphus_2017}%
  \BibitemOpen
  \bibfield  {author} {\bibinfo {author} {\bibfnamefont {I.}~\bibnamefont
  {Kozyryev}}, \bibinfo {author} {\bibfnamefont {L.}~\bibnamefont {Baum}},
  \bibinfo {author} {\bibfnamefont {K.}~\bibnamefont {Matsuda}}, \bibinfo
  {author} {\bibfnamefont {B.~L.}\ \bibnamefont {Augenbraun}}, \bibinfo
  {author} {\bibfnamefont {L.}~\bibnamefont {Anderegg}}, \bibinfo {author}
  {\bibfnamefont {A.~P.}\ \bibnamefont {Sedlack}},\ and\ \bibinfo {author}
  {\bibfnamefont {J.~M.}\ \bibnamefont {Doyle}},\ }\bibfield  {title} {\bibinfo
  {title} {Sisyphus {Laser} {Cooling} of a {Polyatomic} {Molecule}},\ }\href
  {https://doi.org/10.1103/PhysRevLett.118.173201} {\bibfield  {journal}
  {\bibinfo  {journal} {Physical Review Letters}\ }\textbf {\bibinfo {volume}
  {118}},\ \bibinfo {pages} {173201} (\bibinfo {year} {2017})},\ \bibinfo
  {note} {publisher: American Physical Society}\BibitemShut {NoStop}%
\bibitem [{\citenamefont {Mitra}\ \emph {et~al.}(2020)\citenamefont {Mitra},
  \citenamefont {Vilas}, \citenamefont {Hallas}, \citenamefont {Anderegg},
  \citenamefont {Augenbraun}, \citenamefont {Baum}, \citenamefont {Miller},
  \citenamefont {Raval},\ and\ \citenamefont {Doyle}}]{mitra_direct_2020}%
  \BibitemOpen
  \bibfield  {author} {\bibinfo {author} {\bibfnamefont {D.}~\bibnamefont
  {Mitra}}, \bibinfo {author} {\bibfnamefont {N.~B.}\ \bibnamefont {Vilas}},
  \bibinfo {author} {\bibfnamefont {C.}~\bibnamefont {Hallas}}, \bibinfo
  {author} {\bibfnamefont {L.}~\bibnamefont {Anderegg}}, \bibinfo {author}
  {\bibfnamefont {B.~L.}\ \bibnamefont {Augenbraun}}, \bibinfo {author}
  {\bibfnamefont {L.}~\bibnamefont {Baum}}, \bibinfo {author} {\bibfnamefont
  {C.}~\bibnamefont {Miller}}, \bibinfo {author} {\bibfnamefont
  {S.}~\bibnamefont {Raval}},\ and\ \bibinfo {author} {\bibfnamefont {J.~M.}\
  \bibnamefont {Doyle}},\ }\bibfield  {title} {\bibinfo {title} {Direct laser
  cooling of a symmetric top molecule},\ }\href
  {https://doi.org/10.1126/science.abc5357} {\bibfield  {journal} {\bibinfo
  {journal} {Science}\ }\textbf {\bibinfo {volume} {369}},\ \bibinfo {pages}
  {1366} (\bibinfo {year} {2020})},\ \bibinfo {note} {publisher: American
  Association for the Advancement of Science}\BibitemShut {NoStop}%
\bibitem [{\citenamefont {Stollenwerk}\ \emph {et~al.}(2020)\citenamefont
  {Stollenwerk}, \citenamefont {Antonov}, \citenamefont {Venkataramanababu},
  \citenamefont {Lin},\ and\ \citenamefont {Odom}}]{stollenwerk_cooling_2020}%
  \BibitemOpen
  \bibfield  {author} {\bibinfo {author} {\bibfnamefont {P.~R.}\ \bibnamefont
  {Stollenwerk}}, \bibinfo {author} {\bibfnamefont {I.~O.}\ \bibnamefont
  {Antonov}}, \bibinfo {author} {\bibfnamefont {S.}~\bibnamefont
  {Venkataramanababu}}, \bibinfo {author} {\bibfnamefont {Y.-W.}\ \bibnamefont
  {Lin}},\ and\ \bibinfo {author} {\bibfnamefont {B.~C.}\ \bibnamefont
  {Odom}},\ }\bibfield  {title} {\bibinfo {title} {Cooling of a
  {Zero}-{Nuclear}-{Spin} {Molecular} {Ion} to a {Selected} {Rotational}
  {State}},\ }\href {https://doi.org/10.1103/PhysRevLett.125.113201} {\bibfield
   {journal} {\bibinfo  {journal} {Physical Review Letters}\ }\textbf {\bibinfo
  {volume} {125}},\ \bibinfo {pages} {113201} (\bibinfo {year} {2020})},\
  \bibinfo {note} {publisher: American Physical Society}\BibitemShut {NoStop}%
\bibitem [{\citenamefont {Mitra}\ \emph {et~al.}(2022)\citenamefont {Mitra},
  \citenamefont {Lasner}, \citenamefont {Zhu}, \citenamefont {Dickerson},
  \citenamefont {Augenbraun}, \citenamefont {Bailey}, \citenamefont
  {Alexandrova}, \citenamefont {Campbell}, \citenamefont {Caram}, \citenamefont
  {Hudson},\ and\ \citenamefont {Doyle}}]{mitra_pathway_2022}%
  \BibitemOpen
  \bibfield  {author} {\bibinfo {author} {\bibfnamefont {D.}~\bibnamefont
  {Mitra}}, \bibinfo {author} {\bibfnamefont {Z.~D.}\ \bibnamefont {Lasner}},
  \bibinfo {author} {\bibfnamefont {G.-Z.}\ \bibnamefont {Zhu}}, \bibinfo
  {author} {\bibfnamefont {C.~E.}\ \bibnamefont {Dickerson}}, \bibinfo {author}
  {\bibfnamefont {B.~L.}\ \bibnamefont {Augenbraun}}, \bibinfo {author}
  {\bibfnamefont {A.~D.}\ \bibnamefont {Bailey}}, \bibinfo {author}
  {\bibfnamefont {A.~N.}\ \bibnamefont {Alexandrova}}, \bibinfo {author}
  {\bibfnamefont {W.~C.}\ \bibnamefont {Campbell}}, \bibinfo {author}
  {\bibfnamefont {J.~R.}\ \bibnamefont {Caram}}, \bibinfo {author}
  {\bibfnamefont {E.~R.}\ \bibnamefont {Hudson}},\ and\ \bibinfo {author}
  {\bibfnamefont {J.~M.}\ \bibnamefont {Doyle}},\ }\bibfield  {title} {\bibinfo
  {title} {Pathway toward {Optical} {Cycling} and {Laser} {Cooling} of
  {Functionalized} {Arenes}},\ }\href
  {https://doi.org/10.1021/acs.jpclett.2c01430} {\bibfield  {journal} {\bibinfo
   {journal} {The Journal of Physical Chemistry Letters}\ }\textbf {\bibinfo
  {volume} {13}},\ \bibinfo {pages} {7029} (\bibinfo {year} {2022})},\ \bibinfo
  {note} {publisher: American Chemical Society}\BibitemShut {NoStop}%
\bibitem [{\citenamefont {Sawaoka}\ \emph {et~al.}(2023)\citenamefont
  {Sawaoka}, \citenamefont {Frenett}, \citenamefont {Nasir}, \citenamefont
  {Ono}, \citenamefont {Augenbraun}, \citenamefont {Steimle},\ and\
  \citenamefont {Doyle}}]{sawaoka_zeeman-sisyphus_2023}%
  \BibitemOpen
  \bibfield  {author} {\bibinfo {author} {\bibfnamefont {H.}~\bibnamefont
  {Sawaoka}}, \bibinfo {author} {\bibfnamefont {A.}~\bibnamefont {Frenett}},
  \bibinfo {author} {\bibfnamefont {A.}~\bibnamefont {Nasir}}, \bibinfo
  {author} {\bibfnamefont {T.}~\bibnamefont {Ono}}, \bibinfo {author}
  {\bibfnamefont {B.~L.}\ \bibnamefont {Augenbraun}}, \bibinfo {author}
  {\bibfnamefont {T.~C.}\ \bibnamefont {Steimle}},\ and\ \bibinfo {author}
  {\bibfnamefont {J.~M.}\ \bibnamefont {Doyle}},\ }\bibfield  {title} {\bibinfo
  {title} {Zeeman-{Sisyphus} deceleration for heavy molecules with perturbed
  excited-state structure},\ }\href
  {https://doi.org/10.1103/PhysRevA.107.022810} {\bibfield  {journal} {\bibinfo
   {journal} {Physical Review A}\ }\textbf {\bibinfo {volume} {107}},\ \bibinfo
  {pages} {022810} (\bibinfo {year} {2023})},\ \bibinfo {note} {publisher:
  American Physical Society}\BibitemShut {NoStop}%
\bibitem [{\citenamefont {Mills}\ \emph {et~al.}(2020)\citenamefont {Mills},
  \citenamefont {Wu}, \citenamefont {Reed}, \citenamefont {Qi}, \citenamefont
  {Brown}, \citenamefont {Schneider}, \citenamefont {Heaven}, \citenamefont
  {Campbell},\ and\ \citenamefont {Hudson}}]{mills_dipolephonon_2020}%
  \BibitemOpen
  \bibfield  {author} {\bibinfo {author} {\bibfnamefont {M.}~\bibnamefont
  {Mills}}, \bibinfo {author} {\bibfnamefont {H.}~\bibnamefont {Wu}}, \bibinfo
  {author} {\bibfnamefont {E.~C.}\ \bibnamefont {Reed}}, \bibinfo {author}
  {\bibfnamefont {L.}~\bibnamefont {Qi}}, \bibinfo {author} {\bibfnamefont
  {K.~R.}\ \bibnamefont {Brown}}, \bibinfo {author} {\bibfnamefont
  {C.}~\bibnamefont {Schneider}}, \bibinfo {author} {\bibfnamefont {M.~C.}\
  \bibnamefont {Heaven}}, \bibinfo {author} {\bibfnamefont {W.~C.}\
  \bibnamefont {Campbell}},\ and\ \bibinfo {author} {\bibfnamefont {E.~R.}\
  \bibnamefont {Hudson}},\ }\bibfield  {title} {\bibinfo {title}
  {Dipole–phonon quantum logic with alkaline-earth monoxide and monosulfide
  cations},\ }\href {https://doi.org/10.1039/D0CP04574H} {\bibfield  {journal}
  {\bibinfo  {journal} {Physical Chemistry Chemical Physics}\ }\textbf
  {\bibinfo {volume} {22}},\ \bibinfo {pages} {24964} (\bibinfo {year}
  {2020})},\ \bibinfo {note} {publisher: The Royal Society of
  Chemistry}\BibitemShut {NoStop}%
\bibitem [{\citenamefont {Campbell}\ and\ \citenamefont
  {Hudson}(2019)}]{campbell_dipole-phonon_2019}%
  \BibitemOpen
  \bibfield  {author} {\bibinfo {author} {\bibfnamefont {W.~C.}\ \bibnamefont
  {Campbell}}\ and\ \bibinfo {author} {\bibfnamefont {E.~R.}\ \bibnamefont
  {Hudson}},\ }\bibfield  {title} {\bibinfo {title} {Dipole-phonon quantum
  logic with trapped polar molecular ions},\ }\href
  {http://arxiv.org/abs/1909.02668} {\bibfield  {journal} {\bibinfo  {journal}
  {e-print}\ } (\bibinfo {year} {2019})}\BibitemShut {NoStop}%
\bibitem [{\citenamefont {Mintert}\ and\ \citenamefont
  {Wunderlich}(2001)}]{mintert_ion-trap_2001}%
  \BibitemOpen
  \bibfield  {author} {\bibinfo {author} {\bibfnamefont {F.}~\bibnamefont
  {Mintert}}\ and\ \bibinfo {author} {\bibfnamefont {C.}~\bibnamefont
  {Wunderlich}},\ }\bibfield  {title} {\bibinfo {title} {Ion-{Trap} {Quantum}
  {Logic} {Using} {Long}-{Wavelength} {Radiation}},\ }\href
  {https://doi.org/10.1103/PhysRevLett.87.257904} {\bibfield  {journal}
  {\bibinfo  {journal} {Physical Review Letters}\ }\textbf {\bibinfo {volume}
  {87}},\ \bibinfo {pages} {257904} (\bibinfo {year} {2001})},\ \bibinfo {note}
  {publisher: American Physical Society}\BibitemShut {NoStop}%
\bibitem [{\citenamefont {Johanning}\ \emph {et~al.}(2009)\citenamefont
  {Johanning}, \citenamefont {Braun}, \citenamefont {Timoney}, \citenamefont
  {Elman}, \citenamefont {Neuhauser},\ and\ \citenamefont
  {Wunderlich}}]{johanning_individual_2009}%
  \BibitemOpen
  \bibfield  {author} {\bibinfo {author} {\bibfnamefont {M.}~\bibnamefont
  {Johanning}}, \bibinfo {author} {\bibfnamefont {A.}~\bibnamefont {Braun}},
  \bibinfo {author} {\bibfnamefont {N.}~\bibnamefont {Timoney}}, \bibinfo
  {author} {\bibfnamefont {V.}~\bibnamefont {Elman}}, \bibinfo {author}
  {\bibfnamefont {W.}~\bibnamefont {Neuhauser}},\ and\ \bibinfo {author}
  {\bibfnamefont {C.}~\bibnamefont {Wunderlich}},\ }\bibfield  {title}
  {\bibinfo {title} {Individual {Addressing} of {Trapped} {Ions} and {Coupling}
  of {Motional} and {Spin} {States} {Using} rf {Radiation}},\ }\href
  {https://doi.org/10.1103/PhysRevLett.102.073004} {\bibfield  {journal}
  {\bibinfo  {journal} {Physical Review Letters}\ }\textbf {\bibinfo {volume}
  {102}},\ \bibinfo {pages} {073004} (\bibinfo {year} {2009})},\ \bibinfo
  {note} {publisher: American Physical Society}\BibitemShut {NoStop}%
\bibitem [{\citenamefont {Ospelkaus}\ \emph {et~al.}(2011)\citenamefont
  {Ospelkaus}, \citenamefont {Warring}, \citenamefont {Colombe}, \citenamefont
  {Brown}, \citenamefont {Amini}, \citenamefont {Leibfried},\ and\
  \citenamefont {Wineland}}]{ospelkaus_microwave_2011}%
  \BibitemOpen
  \bibfield  {author} {\bibinfo {author} {\bibfnamefont {C.}~\bibnamefont
  {Ospelkaus}}, \bibinfo {author} {\bibfnamefont {U.}~\bibnamefont {Warring}},
  \bibinfo {author} {\bibfnamefont {Y.}~\bibnamefont {Colombe}}, \bibinfo
  {author} {\bibfnamefont {K.~R.}\ \bibnamefont {Brown}}, \bibinfo {author}
  {\bibfnamefont {J.~M.}\ \bibnamefont {Amini}}, \bibinfo {author}
  {\bibfnamefont {D.}~\bibnamefont {Leibfried}},\ and\ \bibinfo {author}
  {\bibfnamefont {D.~J.}\ \bibnamefont {Wineland}},\ }\bibfield  {title}
  {\bibinfo {title} {Microwave quantum logic gates for trapped ions},\ }\href
  {https://doi.org/10.1038/nature10290} {\bibfield  {journal} {\bibinfo
  {journal} {Nature}\ }\textbf {\bibinfo {volume} {476}},\ \bibinfo {pages}
  {181} (\bibinfo {year} {2011})},\ \bibinfo {note} {arXiv:1104.3573 [physics,
  physics:quant-ph]}\BibitemShut {NoStop}%
\bibitem [{\citenamefont {Brown}\ \emph {et~al.}(2011)\citenamefont {Brown},
  \citenamefont {Wilson}, \citenamefont {Colombe}, \citenamefont {Ospelkaus},
  \citenamefont {Meier}, \citenamefont {Knill}, \citenamefont {Leibfried},\
  and\ \citenamefont {Wineland}}]{brown_single-qubit-gate_2011}%
  \BibitemOpen
  \bibfield  {author} {\bibinfo {author} {\bibfnamefont {K.~R.}\ \bibnamefont
  {Brown}}, \bibinfo {author} {\bibfnamefont {A.~C.}\ \bibnamefont {Wilson}},
  \bibinfo {author} {\bibfnamefont {Y.}~\bibnamefont {Colombe}}, \bibinfo
  {author} {\bibfnamefont {C.}~\bibnamefont {Ospelkaus}}, \bibinfo {author}
  {\bibfnamefont {A.~M.}\ \bibnamefont {Meier}}, \bibinfo {author}
  {\bibfnamefont {E.}~\bibnamefont {Knill}}, \bibinfo {author} {\bibfnamefont
  {D.}~\bibnamefont {Leibfried}},\ and\ \bibinfo {author} {\bibfnamefont
  {D.~J.}\ \bibnamefont {Wineland}},\ }\bibfield  {title} {\bibinfo {title}
  {Single-qubit-gate error below $\mathbf{10^{-4}}$ in a trapped ion},\ }\href
  {https://doi.org/10.1103/PhysRevA.84.030303} {\bibfield  {journal} {\bibinfo
  {journal} {Physical Review A}\ }\textbf {\bibinfo {volume} {84}},\ \bibinfo
  {pages} {030303(R)} (\bibinfo {year} {2011})},\ \bibinfo {note} {publisher:
  American Physical Society}\BibitemShut {NoStop}%
\bibitem [{\citenamefont {Timoney}\ \emph {et~al.}(2011)\citenamefont
  {Timoney}, \citenamefont {Baumgart}, \citenamefont {Johanning}, \citenamefont
  {Varon}, \citenamefont {Wunderlich}, \citenamefont {Plenio},\ and\
  \citenamefont {Retzker}}]{timoney_quantum_2011}%
  \BibitemOpen
  \bibfield  {author} {\bibinfo {author} {\bibfnamefont {N.}~\bibnamefont
  {Timoney}}, \bibinfo {author} {\bibfnamefont {I.}~\bibnamefont {Baumgart}},
  \bibinfo {author} {\bibfnamefont {M.}~\bibnamefont {Johanning}}, \bibinfo
  {author} {\bibfnamefont {A.~F.}\ \bibnamefont {Varon}}, \bibinfo {author}
  {\bibfnamefont {C.}~\bibnamefont {Wunderlich}}, \bibinfo {author}
  {\bibfnamefont {M.~B.}\ \bibnamefont {Plenio}},\ and\ \bibinfo {author}
  {\bibfnamefont {A.}~\bibnamefont {Retzker}},\ }\bibfield  {title} {\bibinfo
  {title} {Quantum {Gates} and {Memory} using {Microwave} {Dressed} {States}},\
  }\href {https://doi.org/10.1038/nature10319} {\bibfield  {journal} {\bibinfo
  {journal} {Nature}\ }\textbf {\bibinfo {volume} {476}},\ \bibinfo {pages}
  {185} (\bibinfo {year} {2011})},\ \bibinfo {note} {arXiv:1105.1146
  [quant-ph]}\BibitemShut {NoStop}%
\bibitem [{\citenamefont {Sutherland}\ \emph {et~al.}(2019)\citenamefont
  {Sutherland}, \citenamefont {Srinivas}, \citenamefont {Burd}, \citenamefont
  {Leibfried}, \citenamefont {Wilson}, \citenamefont {Wineland}, \citenamefont
  {Allcock}, \citenamefont {Slichter},\ and\ \citenamefont
  {Libby}}]{sutherland_versatile_2019}%
  \BibitemOpen
  \bibfield  {author} {\bibinfo {author} {\bibfnamefont {R.~T.}\ \bibnamefont
  {Sutherland}}, \bibinfo {author} {\bibfnamefont {R.}~\bibnamefont
  {Srinivas}}, \bibinfo {author} {\bibfnamefont {S.~C.}\ \bibnamefont {Burd}},
  \bibinfo {author} {\bibfnamefont {D.}~\bibnamefont {Leibfried}}, \bibinfo
  {author} {\bibfnamefont {A.~C.}\ \bibnamefont {Wilson}}, \bibinfo {author}
  {\bibfnamefont {D.~J.}\ \bibnamefont {Wineland}}, \bibinfo {author}
  {\bibfnamefont {D.~T.~C.}\ \bibnamefont {Allcock}}, \bibinfo {author}
  {\bibfnamefont {D.~H.}\ \bibnamefont {Slichter}},\ and\ \bibinfo {author}
  {\bibfnamefont {S.~B.}\ \bibnamefont {Libby}},\ }\bibfield  {title} {\bibinfo
  {title} {Versatile laser-free trapped-ion entangling gates},\ }\href
  {https://doi.org/10.1088/1367-2630/ab0be5} {\bibfield  {journal} {\bibinfo
  {journal} {New Journal of Physics}\ }\textbf {\bibinfo {volume} {21}},\
  \bibinfo {pages} {033033} (\bibinfo {year} {2019})},\ \bibinfo {note}
  {publisher: IOP Publishing}\BibitemShut {NoStop}%
\bibitem [{\citenamefont {DeMille}(2002)}]{demille_quantum_2002}%
  \BibitemOpen
  \bibfield  {author} {\bibinfo {author} {\bibfnamefont {D.}~\bibnamefont
  {DeMille}},\ }\bibfield  {title} {\bibinfo {title} {Quantum {Computation}
  with {Trapped} {Polar} {Molecules}},\ }\href
  {https://doi.org/10.1103/PhysRevLett.88.067901} {\bibfield  {journal}
  {\bibinfo  {journal} {Phys. Rev. Lett.}\ }\textbf {\bibinfo {volume} {88}},\
  \bibinfo {pages} {067901} (\bibinfo {year} {2002})}\BibitemShut {NoStop}%
\bibitem [{\citenamefont {Zeppenfeld}(2023)}]{zeppenfeld2023robust}%
  \BibitemOpen
  \bibfield  {author} {\bibinfo {author} {\bibfnamefont {M.}~\bibnamefont
  {Zeppenfeld}},\ }\bibfield  {title} {\bibinfo {title} {A robust framework for
  quantum computation using quasi-hidden molecular degrees of freedom},\
  }\href@noop {} {\bibfield  {journal} {\bibinfo  {journal} {arXiv preprint
  arXiv:2311.14133}\ } (\bibinfo {year} {2023})}\BibitemShut {NoStop}%
\bibitem [{\citenamefont {Wei}\ \emph {et~al.}(2016)\citenamefont {Wei},
  \citenamefont {Cao}, \citenamefont {Kais}, \citenamefont {Friedrich},\ and\
  \citenamefont {Herschbach}}]{wei_quantum_2016}%
  \BibitemOpen
  \bibfield  {author} {\bibinfo {author} {\bibfnamefont {Q.}~\bibnamefont
  {Wei}}, \bibinfo {author} {\bibfnamefont {Y.}~\bibnamefont {Cao}}, \bibinfo
  {author} {\bibfnamefont {S.}~\bibnamefont {Kais}}, \bibinfo {author}
  {\bibfnamefont {B.}~\bibnamefont {Friedrich}},\ and\ \bibinfo {author}
  {\bibfnamefont {D.}~\bibnamefont {Herschbach}},\ }\bibfield  {title}
  {\bibinfo {title} {Quantum {Computation} using {Arrays} of {N} {Polar}
  {Molecules} in {Pendular} {States}},\ }\href
  {https://doi.org/10.1002/cphc.201600781} {\bibfield  {journal} {\bibinfo
  {journal} {Chemphyschem: A European Journal of Chemical Physics and Physical
  Chemistry}\ }\textbf {\bibinfo {volume} {17}},\ \bibinfo {pages} {3714}
  (\bibinfo {year} {2016})}\BibitemShut {NoStop}%
\bibitem [{\citenamefont {Albert}\ \emph {et~al.}(2020)\citenamefont {Albert},
  \citenamefont {Covey},\ and\ \citenamefont {Preskill}}]{albert_robust_2020}%
  \BibitemOpen
  \bibfield  {author} {\bibinfo {author} {\bibfnamefont {V.~V.}\ \bibnamefont
  {Albert}}, \bibinfo {author} {\bibfnamefont {J.~P.}\ \bibnamefont {Covey}},\
  and\ \bibinfo {author} {\bibfnamefont {J.}~\bibnamefont {Preskill}},\
  }\bibfield  {title} {\bibinfo {title} {Robust encoding of a qubit in a
  molecule},\ }\href {https://doi.org/10.1103/PhysRevX.10.031050} {\bibfield
  {journal} {\bibinfo  {journal} {Physical Review X}\ }\textbf {\bibinfo
  {volume} {10}},\ \bibinfo {pages} {031050} (\bibinfo {year} {2020})},\
  \bibinfo {note} {arXiv:1911.00099 [cond-mat, physics:physics,
  physics:quant-ph]}\BibitemShut {NoStop}%
\bibitem [{\citenamefont {Hollas}(2004)}]{hollas2004modern}%
  \BibitemOpen
  \bibfield  {author} {\bibinfo {author} {\bibfnamefont {J.~M.}\ \bibnamefont
  {Hollas}},\ }\href@noop {} {\emph {\bibinfo {title} {Modern spectroscopy}}}\
  (\bibinfo  {publisher} {John Wiley \& Sons},\ \bibinfo {year}
  {2004})\BibitemShut {NoStop}%
\bibitem [{\citenamefont {Cohen-Tannoudji}\ \emph {et~al.}(1998)\citenamefont
  {Cohen-Tannoudji}, \citenamefont {Dupont-Roc},\ and\ \citenamefont
  {Grynberg}}]{cohen-tannoudji_atom-photon_1998}%
  \BibitemOpen
  \bibfield  {author} {\bibinfo {author} {\bibfnamefont {C.}~\bibnamefont
  {Cohen-Tannoudji}}, \bibinfo {author} {\bibfnamefont {J.}~\bibnamefont
  {Dupont-Roc}},\ and\ \bibinfo {author} {\bibfnamefont {G.}~\bibnamefont
  {Grynberg}},\ }\href
  {http://www.wiley.com/WileyCDA/WileyTitle/productCd-0471293369.html} {\emph
  {\bibinfo {title} {Atom-{Photon} {Interactions}: {Basic} {Processes} and
  {Applications}}}}\ (\bibinfo  {publisher} {Wiley},\ \bibinfo {address}
  {Morlenbach},\ \bibinfo {year} {1998})\BibitemShut {NoStop}%
\bibitem [{\citenamefont {Nielsen}\ and\ \citenamefont
  {Chuang}(2011)}]{nielsen_quantum_2011}%
  \BibitemOpen
  \bibfield  {author} {\bibinfo {author} {\bibfnamefont {M.~A.}\ \bibnamefont
  {Nielsen}}\ and\ \bibinfo {author} {\bibfnamefont {I.~L.}\ \bibnamefont
  {Chuang}},\ }\href@noop {} {\emph {\bibinfo {title} {Quantum {Computation}
  and {Quantum} {Information}: 10th {Anniversary} {Edition}}}},\ \bibinfo
  {edition} {10th}\ ed.\ (\bibinfo  {publisher} {Cambridge University Press},\
  \bibinfo {address} {New York, NY, USA},\ \bibinfo {year} {2011})\BibitemShut
  {NoStop}%
\bibitem [{\citenamefont {Gardiner}\ and\ \citenamefont
  {Zoller}(2015)}]{gardiner_quantum_2015}%
  \BibitemOpen
  \bibfield  {author} {\bibinfo {author} {\bibfnamefont {C.}~\bibnamefont
  {Gardiner}}\ and\ \bibinfo {author} {\bibfnamefont {P.}~\bibnamefont
  {Zoller}},\ }\href@noop {} {\emph {\bibinfo {title} {The quantum world of
  ultra-cold atoms and light. {Book} {II}}}},\ edited by\ \bibinfo {editor}
  {\bibfnamefont {C.}~\bibnamefont {Salomon}}\ (\bibinfo  {publisher} {Imperial
  College Press},\ \bibinfo {year} {2015})\BibitemShut {NoStop}%
\bibitem [{\citenamefont {Breuer}(2007)}]{breuer_theory_2007}%
  \BibitemOpen
  \bibfield  {author} {\bibinfo {author} {\bibfnamefont {H.-P.}\ \bibnamefont
  {Breuer}},\ }\href@noop {} {\emph {\bibinfo {title} {The {Theory} of {Open}
  {Quantum} {Systems}}}}\ (\bibinfo  {publisher} {Oxford University Press,
  USA},\ \bibinfo {address} {Oxford},\ \bibinfo {year} {2007})\BibitemShut
  {NoStop}%
\bibitem [{\citenamefont {Albert}\ \emph {et~al.}(2018)\citenamefont {Albert},
  \citenamefont {Noh}, \citenamefont {Duivenvoorden}, \citenamefont {Young},
  \citenamefont {Brierley}, \citenamefont {Reinhold}, \citenamefont {Vuillot},
  \citenamefont {Li}, \citenamefont {Shen}, \citenamefont {Girvin},
  \citenamefont {Terhal},\ and\ \citenamefont
  {Jiang}}]{albert_performance_2018}%
  \BibitemOpen
  \bibfield  {author} {\bibinfo {author} {\bibfnamefont {V.~V.}\ \bibnamefont
  {Albert}}, \bibinfo {author} {\bibfnamefont {K.}~\bibnamefont {Noh}},
  \bibinfo {author} {\bibfnamefont {K.}~\bibnamefont {Duivenvoorden}}, \bibinfo
  {author} {\bibfnamefont {D.~J.}\ \bibnamefont {Young}}, \bibinfo {author}
  {\bibfnamefont {R.~T.}\ \bibnamefont {Brierley}}, \bibinfo {author}
  {\bibfnamefont {P.}~\bibnamefont {Reinhold}}, \bibinfo {author}
  {\bibfnamefont {C.}~\bibnamefont {Vuillot}}, \bibinfo {author} {\bibfnamefont
  {L.}~\bibnamefont {Li}}, \bibinfo {author} {\bibfnamefont {C.}~\bibnamefont
  {Shen}}, \bibinfo {author} {\bibfnamefont {S.~M.}\ \bibnamefont {Girvin}},
  \bibinfo {author} {\bibfnamefont {B.~M.}\ \bibnamefont {Terhal}},\ and\
  \bibinfo {author} {\bibfnamefont {L.}~\bibnamefont {Jiang}},\ }\bibfield
  {title} {\bibinfo {title} {Performance and structure of single-mode bosonic
  codes},\ }\href {https://doi.org/10.1103/PhysRevA.97.032346} {\bibfield
  {journal} {\bibinfo  {journal} {Phys. Rev. A}\ }\textbf {\bibinfo {volume}
  {97}},\ \bibinfo {pages} {032346} (\bibinfo {year} {2018})}\BibitemShut
  {NoStop}%
\bibitem [{\citenamefont {Wineland}\ \emph {et~al.}(2003)\citenamefont
  {Wineland}, \citenamefont {Barrett}, \citenamefont {Britton}, \citenamefont
  {Chiaverini}, \citenamefont {DeMarco}, \citenamefont {Itano}, \citenamefont
  {Jelenkovi'c}, \citenamefont {Langer}, \citenamefont {Leibfried},
  \citenamefont {Meyer}, \citenamefont {Rosenband},\ and\ \citenamefont
  {Schätz}}]{wineland_quantum_2003}%
  \BibitemOpen
  \bibfield  {author} {\bibinfo {author} {\bibfnamefont {D.~J.}\ \bibnamefont
  {Wineland}}, \bibinfo {author} {\bibfnamefont {M.}~\bibnamefont {Barrett}},
  \bibinfo {author} {\bibfnamefont {J.}~\bibnamefont {Britton}}, \bibinfo
  {author} {\bibfnamefont {J.}~\bibnamefont {Chiaverini}}, \bibinfo {author}
  {\bibfnamefont {B.}~\bibnamefont {DeMarco}}, \bibinfo {author} {\bibfnamefont
  {W.~M.}\ \bibnamefont {Itano}}, \bibinfo {author} {\bibfnamefont
  {B.}~\bibnamefont {Jelenkovi'c}}, \bibinfo {author} {\bibfnamefont
  {C.}~\bibnamefont {Langer}}, \bibinfo {author} {\bibfnamefont
  {D.}~\bibnamefont {Leibfried}}, \bibinfo {author} {\bibfnamefont
  {V.}~\bibnamefont {Meyer}}, \bibinfo {author} {\bibfnamefont
  {T.}~\bibnamefont {Rosenband}},\ and\ \bibinfo {author} {\bibfnamefont
  {T.}~\bibnamefont {Schätz}},\ }\bibfield  {title} {\bibinfo {title} {Quantum
  information processing with trapped ions},\ }\href
  {https://doi.org/10.1098/rsta.2003.1205} {\bibfield  {journal} {\bibinfo
  {journal} {Philosophical Transactions of the Royal Society of London. Series
  A: Mathematical, Physical and Engineering Sciences}\ }\textbf {\bibinfo
  {volume} {361}},\ \bibinfo {pages} {1349} (\bibinfo {year} {2003})},\
  \bibinfo {note} {arXiv:quant-ph/0212079}\BibitemShut {NoStop}%
\bibitem [{\citenamefont {Ozeri}\ \emph {et~al.}(2007)\citenamefont {Ozeri},
  \citenamefont {Itano}, \citenamefont {Blakestad}, \citenamefont {Britton},
  \citenamefont {Chiaverini}, \citenamefont {Jost}, \citenamefont {Langer},
  \citenamefont {Leibfried}, \citenamefont {Reichle}, \citenamefont {Seidelin},
  \citenamefont {Wesenberg},\ and\ \citenamefont
  {Wineland}}]{ozeri_errors_2007}%
  \BibitemOpen
  \bibfield  {author} {\bibinfo {author} {\bibfnamefont {R.}~\bibnamefont
  {Ozeri}}, \bibinfo {author} {\bibfnamefont {W.~M.}\ \bibnamefont {Itano}},
  \bibinfo {author} {\bibfnamefont {R.~B.}\ \bibnamefont {Blakestad}}, \bibinfo
  {author} {\bibfnamefont {J.}~\bibnamefont {Britton}}, \bibinfo {author}
  {\bibfnamefont {J.}~\bibnamefont {Chiaverini}}, \bibinfo {author}
  {\bibfnamefont {J.~D.}\ \bibnamefont {Jost}}, \bibinfo {author}
  {\bibfnamefont {C.}~\bibnamefont {Langer}}, \bibinfo {author} {\bibfnamefont
  {D.}~\bibnamefont {Leibfried}}, \bibinfo {author} {\bibfnamefont
  {R.}~\bibnamefont {Reichle}}, \bibinfo {author} {\bibfnamefont
  {S.}~\bibnamefont {Seidelin}}, \bibinfo {author} {\bibfnamefont {J.~H.}\
  \bibnamefont {Wesenberg}},\ and\ \bibinfo {author} {\bibfnamefont {D.~J.}\
  \bibnamefont {Wineland}},\ }\bibfield  {title} {\bibinfo {title} {Errors in
  trapped-ion quantum gates due to spontaneous photon scattering},\ }\href
  {https://doi.org/10.1103/PhysRevA.75.042329} {\bibfield  {journal} {\bibinfo
  {journal} {Physical Review A}\ }\textbf {\bibinfo {volume} {75}},\ \bibinfo
  {pages} {042329} (\bibinfo {year} {2007})},\ \bibinfo {note} {publisher:
  American Physical Society}\BibitemShut {NoStop}%
\bibitem [{\citenamefont {Knill}\ and\ \citenamefont
  {Laflamme}(1997)}]{knill_theory_1997}%
  \BibitemOpen
  \bibfield  {author} {\bibinfo {author} {\bibfnamefont {E.}~\bibnamefont
  {Knill}}\ and\ \bibinfo {author} {\bibfnamefont {R.}~\bibnamefont
  {Laflamme}},\ }\bibfield  {title} {\bibinfo {title} {Theory of quantum
  error-correcting codes},\ }\href {https://doi.org/10.1103/PhysRevA.55.900}
  {\bibfield  {journal} {\bibinfo  {journal} {Phys. Rev. A}\ }\textbf {\bibinfo
  {volume} {55}},\ \bibinfo {pages} {900} (\bibinfo {year} {1997})}\BibitemShut
  {NoStop}%
\bibitem [{\citenamefont {Bennett}\ \emph {et~al.}(1996)\citenamefont
  {Bennett}, \citenamefont {DiVincenzo}, \citenamefont {Smolin},\ and\
  \citenamefont {Wootters}}]{bennett_mixed-state_1996}%
  \BibitemOpen
  \bibfield  {author} {\bibinfo {author} {\bibfnamefont {C.~H.}\ \bibnamefont
  {Bennett}}, \bibinfo {author} {\bibfnamefont {D.~P.}\ \bibnamefont
  {DiVincenzo}}, \bibinfo {author} {\bibfnamefont {J.~A.}\ \bibnamefont
  {Smolin}},\ and\ \bibinfo {author} {\bibfnamefont {W.~K.}\ \bibnamefont
  {Wootters}},\ }\bibfield  {title} {\bibinfo {title} {Mixed-state entanglement
  and quantum error correction},\ }\href
  {https://doi.org/10.1103/PhysRevA.54.3824} {\bibfield  {journal} {\bibinfo
  {journal} {Phys. Rev. A}\ }\textbf {\bibinfo {volume} {54}},\ \bibinfo
  {pages} {3824} (\bibinfo {year} {1996})}\BibitemShut {NoStop}%
\bibitem [{\citenamefont {Chessa}\ and\ \citenamefont
  {Giovannetti}(2021)}]{chessa_quantum_2021}%
  \BibitemOpen
  \bibfield  {author} {\bibinfo {author} {\bibfnamefont {S.}~\bibnamefont
  {Chessa}}\ and\ \bibinfo {author} {\bibfnamefont {V.}~\bibnamefont
  {Giovannetti}},\ }\bibfield  {title} {\bibinfo {title} {Quantum capacity
  analysis of multi-level amplitude damping channels},\ }\href
  {https://doi.org/10.1038/s42005-021-00524-4} {\bibfield  {journal} {\bibinfo
  {journal} {Communications Physics}\ }\textbf {\bibinfo {volume} {4}},\
  \bibinfo {pages} {22} (\bibinfo {year} {2021})},\ \bibinfo {note}
  {arXiv:2008.00477 [quant-ph]}\BibitemShut {NoStop}%
\bibitem [{\citenamefont {Chessa}\ and\ \citenamefont
  {Giovannetti}(2023)}]{chessa_resonant_2023}%
  \BibitemOpen
  \bibfield  {author} {\bibinfo {author} {\bibfnamefont {S.}~\bibnamefont
  {Chessa}}\ and\ \bibinfo {author} {\bibfnamefont {V.}~\bibnamefont
  {Giovannetti}},\ }\bibfield  {title} {\bibinfo {title} {Resonant {Multilevel}
  {Amplitude} {Damping} {Channels}},\ }\href
  {https://doi.org/10.22331/q-2023-01-19-902} {\bibfield  {journal} {\bibinfo
  {journal} {Quantum}\ }\textbf {\bibinfo {volume} {7}},\ \bibinfo {pages}
  {902} (\bibinfo {year} {2023})},\ \bibinfo {note} {publisher: Verein zur
  Förderung des Open Access Publizierens in den
  Quantenwissenschaften}\BibitemShut {NoStop}%
\bibitem [{\citenamefont {Foot}(2005)}]{foot_atomic_2005}%
  \BibitemOpen
  \bibfield  {author} {\bibinfo {author} {\bibfnamefont {C.~J.}\ \bibnamefont
  {Foot}},\ }\href@noop {} {\emph {\bibinfo {title} {Atomic {Physics}}}},\
  Oxford {Master} {Series} in {Physics}\ (\bibinfo  {publisher} {Oxford
  University Press},\ \bibinfo {address} {Oxford, New York},\ \bibinfo {year}
  {2005})\BibitemShut {NoStop}%
\bibitem [{\citenamefont {Varshalovich}\ \emph {et~al.}(1988)\citenamefont
  {Varshalovich}, \citenamefont {Moskalev},\ and\ \citenamefont
  {Khersonskii}}]{varshalovich_quantum_1988}%
  \BibitemOpen
  \bibfield  {author} {\bibinfo {author} {\bibfnamefont {D.~A.}\ \bibnamefont
  {Varshalovich}}, \bibinfo {author} {\bibfnamefont {A.~N.}\ \bibnamefont
  {Moskalev}},\ and\ \bibinfo {author} {\bibfnamefont {V.~K.}\ \bibnamefont
  {Khersonskii}},\ }\href {https://doi.org/10.1142/0270} {\emph {\bibinfo
  {title} {Quantum {Theory} of {Angular} {Momentum}}}}\ (\bibinfo  {publisher}
  {World Scientific},\ \bibinfo {year} {1988})\BibitemShut {NoStop}%
\bibitem [{\citenamefont {Chuang}\ and\ \citenamefont
  {Yamamoto}(1995)}]{chuang_simple_1995}%
  \BibitemOpen
  \bibfield  {author} {\bibinfo {author} {\bibfnamefont {I.~L.}\ \bibnamefont
  {Chuang}}\ and\ \bibinfo {author} {\bibfnamefont {Y.}~\bibnamefont
  {Yamamoto}},\ }\bibfield  {title} {\bibinfo {title} {Simple quantum
  computer},\ }\href {https://doi.org/10.1103/PhysRevA.52.3489} {\bibfield
  {journal} {\bibinfo  {journal} {Phys. Rev. A}\ }\textbf {\bibinfo {volume}
  {52}},\ \bibinfo {pages} {3489} (\bibinfo {year} {1995})}\BibitemShut
  {NoStop}%
\bibitem [{\citenamefont {Michael}\ \emph {et~al.}(2016)\citenamefont
  {Michael}, \citenamefont {Silveri}, \citenamefont {Brierley}, \citenamefont
  {Albert}, \citenamefont {Salmilehto}, \citenamefont {Jiang},\ and\
  \citenamefont {Girvin}}]{michael_new_2016}%
  \BibitemOpen
  \bibfield  {author} {\bibinfo {author} {\bibfnamefont {M.~H.}\ \bibnamefont
  {Michael}}, \bibinfo {author} {\bibfnamefont {M.}~\bibnamefont {Silveri}},
  \bibinfo {author} {\bibfnamefont {R.~T.}\ \bibnamefont {Brierley}}, \bibinfo
  {author} {\bibfnamefont {V.~V.}\ \bibnamefont {Albert}}, \bibinfo {author}
  {\bibfnamefont {J.}~\bibnamefont {Salmilehto}}, \bibinfo {author}
  {\bibfnamefont {L.}~\bibnamefont {Jiang}},\ and\ \bibinfo {author}
  {\bibfnamefont {S.~M.}\ \bibnamefont {Girvin}},\ }\bibfield  {title}
  {\bibinfo {title} {New {Class} of {Quantum} {Error}-{Correcting} {Codes} for
  a {Bosonic} {Mode}},\ }\href {https://doi.org/10.1103/PhysRevX.6.031006}
  {\bibfield  {journal} {\bibinfo  {journal} {Phys. Rev. X}\ }\textbf {\bibinfo
  {volume} {6}},\ \bibinfo {pages} {031006} (\bibinfo {year}
  {2016})}\BibitemShut {NoStop}%
\bibitem [{\citenamefont {Ouyang}(2014)}]{ouyang_permutation-invariant_2014}%
  \BibitemOpen
  \bibfield  {author} {\bibinfo {author} {\bibfnamefont {Y.}~\bibnamefont
  {Ouyang}},\ }\bibfield  {title} {\bibinfo {title} {Permutation-invariant
  quantum codes},\ }\href {https://doi.org/10.1103/PhysRevA.90.062317}
  {\bibfield  {journal} {\bibinfo  {journal} {Phys. Rev. A}\ }\textbf {\bibinfo
  {volume} {90}},\ \bibinfo {pages} {062317} (\bibinfo {year}
  {2014})}\BibitemShut {NoStop}%
\bibitem [{\citenamefont {Gross}(2021)}]{gross_designing_2021}%
  \BibitemOpen
  \bibfield  {author} {\bibinfo {author} {\bibfnamefont {J.~A.}\ \bibnamefont
  {Gross}},\ }\bibfield  {title} {\bibinfo {title} {Designing {Codes} around
  {Interactions}: {The} {Case} of a {Spin}},\ }\href
  {https://doi.org/10.1103/PhysRevLett.127.010504} {\bibfield  {journal}
  {\bibinfo  {journal} {Physical Review Letters}\ }\textbf {\bibinfo {volume}
  {127}},\ \bibinfo {pages} {010504} (\bibinfo {year} {2021})},\ \bibinfo
  {note} {publisher: American Physical Society}\BibitemShut {NoStop}%
\bibitem [{\citenamefont {Gross}\ \emph {et~al.}(2021)\citenamefont {Gross},
  \citenamefont {Godfrin}, \citenamefont {Blais},\ and\ \citenamefont
  {Dupont-Ferrier}}]{gross_hardware-efficient_2021}%
  \BibitemOpen
  \bibfield  {author} {\bibinfo {author} {\bibfnamefont {J.~A.}\ \bibnamefont
  {Gross}}, \bibinfo {author} {\bibfnamefont {C.}~\bibnamefont {Godfrin}},
  \bibinfo {author} {\bibfnamefont {A.}~\bibnamefont {Blais}},\ and\ \bibinfo
  {author} {\bibfnamefont {E.}~\bibnamefont {Dupont-Ferrier}},\ }\href
  {https://doi.org/10.48550/arXiv.2103.08548} {\bibinfo {title}
  {Hardware-efficient error-correcting codes for large nuclear spins}}
  (\bibinfo {year} {2021}),\ \bibinfo {note} {arXiv:2103.08548
  [quant-ph]}\BibitemShut {NoStop}%
\bibitem [{\citenamefont {Fan}\ \emph {et~al.}(2023)\citenamefont {Fan},
  \citenamefont {Fischler},\ and\ \citenamefont
  {Kubischta}}]{fan_quantum_2023}%
  \BibitemOpen
  \bibfield  {author} {\bibinfo {author} {\bibfnamefont {Y.}~\bibnamefont
  {Fan}}, \bibinfo {author} {\bibfnamefont {W.}~\bibnamefont {Fischler}},\ and\
  \bibinfo {author} {\bibfnamefont {E.}~\bibnamefont {Kubischta}},\ }\href
  {https://doi.org/10.1103/PhysRevA.107.032411} {\bibinfo {title} {Quantum
  {Error} {Correction} in the {Lowest} {Landau} {Level}}} (\bibinfo {year}
  {2023}),\ \bibinfo {note} {arXiv:2210.16957 [cond-mat, physics:hep-th,
  physics:quant-ph]}\BibitemShut {NoStop}%
\bibitem [{\citenamefont {Omanakuttan}\ and\ \citenamefont
  {Volkoff}(2023)}]{omanakuttan_spin_2023}%
  \BibitemOpen
  \bibfield  {author} {\bibinfo {author} {\bibfnamefont {S.}~\bibnamefont
  {Omanakuttan}}\ and\ \bibinfo {author} {\bibfnamefont {T.~J.}\ \bibnamefont
  {Volkoff}},\ }\href {http://arxiv.org/abs/2211.05181} {\bibinfo {title} {Spin
  squeezed {GKP} codes for quantum error correction in atomic ensembles}}
  (\bibinfo {year} {2023}),\ \bibinfo {note} {arXiv:2211.05181
  [quant-ph]}\BibitemShut {NoStop}%
\bibitem [{\citenamefont {Omanakuttan}\ and\ \citenamefont
  {Gross}(2023)}]{omanakuttan_multispin_2023}%
  \BibitemOpen
  \bibfield  {author} {\bibinfo {author} {\bibfnamefont {S.}~\bibnamefont
  {Omanakuttan}}\ and\ \bibinfo {author} {\bibfnamefont {J.~A.}\ \bibnamefont
  {Gross}},\ }\href {https://doi.org/10.48550/arXiv.2304.08611} {\bibinfo
  {title} {Multispin {Clifford} codes for angular momentum errors in spin
  systems}} (\bibinfo {year} {2023}),\ \bibinfo {note} {arXiv:2304.08611
  [quant-ph]}\BibitemShut {NoStop}%
\bibitem [{\citenamefont {{Aditya
  Sivakumar}}(2021)}]{aditya_sivakumar_quantum_2021}%
  \BibitemOpen
  \bibfield  {author} {\bibinfo {author} {\bibnamefont {{Aditya Sivakumar}}},\
  }\href@noop {} {\emph {\bibinfo {title} {Quantum {Error} {Correcting} {Codes}
  for {Large} {Spins}}}},\ \bibinfo {type} {Tech. Rep.}\ (\bibinfo
  {institution} {California Institute of Technology},\ \bibinfo {year}
  {2021})\BibitemShut {NoStop}%
\bibitem [{\citenamefont {Boguslawski}\ \emph {et~al.}(2023)\citenamefont
  {Boguslawski}, \citenamefont {Wall}, \citenamefont {Vizvary}, \citenamefont
  {Moore}, \citenamefont {Bareian}, \citenamefont {Allcock}, \citenamefont
  {Wineland}, \citenamefont {Hudson},\ and\ \citenamefont
  {Campbell}}]{boguslawski_raman_2023}%
  \BibitemOpen
  \bibfield  {author} {\bibinfo {author} {\bibfnamefont {M.~J.}\ \bibnamefont
  {Boguslawski}}, \bibinfo {author} {\bibfnamefont {Z.~J.}\ \bibnamefont
  {Wall}}, \bibinfo {author} {\bibfnamefont {S.~R.}\ \bibnamefont {Vizvary}},
  \bibinfo {author} {\bibfnamefont {I.~D.}\ \bibnamefont {Moore}}, \bibinfo
  {author} {\bibfnamefont {M.}~\bibnamefont {Bareian}}, \bibinfo {author}
  {\bibfnamefont {D.~T.}\ \bibnamefont {Allcock}}, \bibinfo {author}
  {\bibfnamefont {D.~J.}\ \bibnamefont {Wineland}}, \bibinfo {author}
  {\bibfnamefont {E.~R.}\ \bibnamefont {Hudson}},\ and\ \bibinfo {author}
  {\bibfnamefont {W.~C.}\ \bibnamefont {Campbell}},\ }\bibfield  {title}
  {\bibinfo {title} {Raman {Scattering} {Errors} in
  {Stimulated}-{Raman}-{Induced} {Logic} {Gates} in $^{133}\text{Ba}^+$},\
  }\href {https://doi.org/10.1103/PhysRevLett.131.063001} {\bibfield  {journal}
  {\bibinfo  {journal} {Physical Review Letters}\ }\textbf {\bibinfo {volume}
  {131}},\ \bibinfo {pages} {063001} (\bibinfo {year} {2023})},\ \bibinfo
  {note} {publisher: American Physical Society}\BibitemShut {NoStop}%
\bibitem [{\citenamefont {Burt}(1995)}]{burt_demonstration_1995}%
  \BibitemOpen
  \bibfield  {author} {\bibinfo {author} {\bibfnamefont {E.~A. E.~A.}\
  \bibnamefont {Burt}},\ }\emph {\bibinfo {title} {Demonstration of trapped
  single laser cooled indium ions}},\ \href
  {https://digital.lib.washington.edu:443/researchworks/handle/1773/9638}
  {\bibinfo {type} {Thesis}} (\bibinfo {year} {1995})\BibitemShut {NoStop}%
\bibitem [{\citenamefont {Kim}\ \emph {et~al.}(2009)\citenamefont {Kim},
  \citenamefont {Haubrich},\ and\ \citenamefont
  {Meschede}}]{kim_efficient_2009}%
  \BibitemOpen
  \bibfield  {author} {\bibinfo {author} {\bibfnamefont {J.-I.}\ \bibnamefont
  {Kim}}, \bibinfo {author} {\bibfnamefont {D.}~\bibnamefont {Haubrich}},\ and\
  \bibinfo {author} {\bibfnamefont {D.}~\bibnamefont {Meschede}},\ }\bibfield
  {title} {\bibinfo {title} {Efficient sub-{Doppler} laser cooling of an
  {Indium} atomic beam},\ }\href {https://doi.org/10.1364/OE.17.021216}
  {\bibfield  {journal} {\bibinfo  {journal} {Optics Express}\ }\textbf
  {\bibinfo {volume} {17}},\ \bibinfo {pages} {21216} (\bibinfo {year}
  {2009})},\ \bibinfo {note} {publisher: Optica Publishing Group}\BibitemShut
  {NoStop}%
\bibitem [{\citenamefont {Arnold}\ \emph {et~al.}(2018)\citenamefont {Arnold},
  \citenamefont {Kaewuam}, \citenamefont {Roy}, \citenamefont {Tan},\ and\
  \citenamefont {Barrett}}]{arnold_blackbody_2018}%
  \BibitemOpen
  \bibfield  {author} {\bibinfo {author} {\bibfnamefont {K.~J.}\ \bibnamefont
  {Arnold}}, \bibinfo {author} {\bibfnamefont {R.}~\bibnamefont {Kaewuam}},
  \bibinfo {author} {\bibfnamefont {A.}~\bibnamefont {Roy}}, \bibinfo {author}
  {\bibfnamefont {T.~R.}\ \bibnamefont {Tan}},\ and\ \bibinfo {author}
  {\bibfnamefont {M.~D.}\ \bibnamefont {Barrett}},\ }\bibfield  {title}
  {\bibinfo {title} {Blackbody radiation shift assessment for a lutetium ion
  clock},\ }\href {https://doi.org/10.1038/s41467-018-04079-x} {\bibfield
  {journal} {\bibinfo  {journal} {Nature Communications}\ }\textbf {\bibinfo
  {volume} {9}},\ \bibinfo {pages} {1650} (\bibinfo {year} {2018})},\ \bibinfo
  {note} {number: 1 Publisher: Nature Publishing Group}\BibitemShut {NoStop}%
\bibitem [{\citenamefont {Yu}\ \emph {et~al.}(2022)\citenamefont {Yu},
  \citenamefont {Mo}, \citenamefont {Lu}, \citenamefont {Tan},\ and\
  \citenamefont {Nicholson}}]{yu_magneto-optical_2022}%
  \BibitemOpen
  \bibfield  {author} {\bibinfo {author} {\bibfnamefont {X.}~\bibnamefont
  {Yu}}, \bibinfo {author} {\bibfnamefont {J.}~\bibnamefont {Mo}}, \bibinfo
  {author} {\bibfnamefont {T.}~\bibnamefont {Lu}}, \bibinfo {author}
  {\bibfnamefont {T.~Y.}\ \bibnamefont {Tan}},\ and\ \bibinfo {author}
  {\bibfnamefont {T.~L.}\ \bibnamefont {Nicholson}},\ }\bibfield  {title}
  {\bibinfo {title} {Magneto-optical trapping of a group-{III} atom},\ }\href
  {https://doi.org/10.1103/PhysRevA.105.L061101} {\bibfield  {journal}
  {\bibinfo  {journal} {Physical Review A}\ }\textbf {\bibinfo {volume}
  {105}},\ \bibinfo {pages} {L061101} (\bibinfo {year} {2022})},\ \bibinfo
  {note} {publisher: American Physical Society}\BibitemShut {NoStop}%
\bibitem [{\citenamefont {Miao}\ \emph {et~al.}(2014)\citenamefont {Miao},
  \citenamefont {Hostetter}, \citenamefont {Stratis},\ and\ \citenamefont
  {Saffman}}]{miao_magneto-optical_2014}%
  \BibitemOpen
  \bibfield  {author} {\bibinfo {author} {\bibfnamefont {J.}~\bibnamefont
  {Miao}}, \bibinfo {author} {\bibfnamefont {J.}~\bibnamefont {Hostetter}},
  \bibinfo {author} {\bibfnamefont {G.}~\bibnamefont {Stratis}},\ and\ \bibinfo
  {author} {\bibfnamefont {M.}~\bibnamefont {Saffman}},\ }\bibfield  {title}
  {\bibinfo {title} {Magneto-optical trapping of holmium atoms},\ }\href
  {https://doi.org/10.1103/PhysRevA.89.041401} {\bibfield  {journal} {\bibinfo
  {journal} {Phys. Rev. A}\ }\textbf {\bibinfo {volume} {89}},\ \bibinfo
  {pages} {41401} (\bibinfo {year} {2014})}\BibitemShut {NoStop}%
\bibitem [{\citenamefont {Lu}\ \emph {et~al.}(2010)\citenamefont {Lu},
  \citenamefont {Youn},\ and\ \citenamefont {Lev}}]{lu_trapping_2010}%
  \BibitemOpen
  \bibfield  {author} {\bibinfo {author} {\bibfnamefont {M.}~\bibnamefont
  {Lu}}, \bibinfo {author} {\bibfnamefont {S.~H.}\ \bibnamefont {Youn}},\ and\
  \bibinfo {author} {\bibfnamefont {B.~L.}\ \bibnamefont {Lev}},\ }\bibfield
  {title} {\bibinfo {title} {Trapping {Ultracold} {Dysprosium}: {A} {Highly}
  {Magnetic} {Gas} for {Dipolar} {Physics}},\ }\href
  {https://doi.org/10.1103/PhysRevLett.104.063001} {\bibfield  {journal}
  {\bibinfo  {journal} {Physical Review Letters}\ }\textbf {\bibinfo {volume}
  {104}},\ \bibinfo {pages} {063001} (\bibinfo {year} {2010})},\ \bibinfo
  {note} {publisher: American Physical Society}\BibitemShut {NoStop}%
\bibitem [{\citenamefont {McClelland}\ and\ \citenamefont
  {Hanssen}(2006)}]{mcclelland_laser_2006}%
  \BibitemOpen
  \bibfield  {author} {\bibinfo {author} {\bibfnamefont {J.}~\bibnamefont
  {McClelland}}\ and\ \bibinfo {author} {\bibfnamefont {J.}~\bibnamefont
  {Hanssen}},\ }\bibfield  {title} {\bibinfo {title} {Laser {Cooling} without
  {Repumping}: {A} {Magneto}-{Optical} {Trap} for {Erbium} {Atoms}},\ }\href
  {https://doi.org/10.1103/PhysRevLett.96.143005} {\bibfield  {journal}
  {\bibinfo  {journal} {Phys. Rev. Lett.}\ }\textbf {\bibinfo {volume} {96}},\
  \bibinfo {pages} {143005} (\bibinfo {year} {2006})}\BibitemShut {NoStop}%
\bibitem [{\citenamefont {Bao}\ \emph {et~al.}(2022)\citenamefont {Bao},
  \citenamefont {Yu}, \citenamefont {Anderegg}, \citenamefont {Chae},
  \citenamefont {Ketterle}, \citenamefont {Ni},\ and\ \citenamefont
  {Doyle}}]{bao_dipolar_2022}%
  \BibitemOpen
  \bibfield  {author} {\bibinfo {author} {\bibfnamefont {Y.}~\bibnamefont
  {Bao}}, \bibinfo {author} {\bibfnamefont {S.~S.}\ \bibnamefont {Yu}},
  \bibinfo {author} {\bibfnamefont {L.}~\bibnamefont {Anderegg}}, \bibinfo
  {author} {\bibfnamefont {E.}~\bibnamefont {Chae}}, \bibinfo {author}
  {\bibfnamefont {W.}~\bibnamefont {Ketterle}}, \bibinfo {author}
  {\bibfnamefont {K.-K.}\ \bibnamefont {Ni}},\ and\ \bibinfo {author}
  {\bibfnamefont {J.~M.}\ \bibnamefont {Doyle}},\ }\href
  {http://arxiv.org/abs/2211.09780} {\bibinfo {title} {Dipolar spin-exchange
  and entanglement between molecules in an optical tweezer array}} (\bibinfo
  {year} {2022}),\ \bibinfo {note} {arXiv:2211.09780 [physics,
  physics:quant-ph]}\BibitemShut {NoStop}%
\bibitem [{\citenamefont {Zhu}\ \emph {et~al.}(2022)\citenamefont {Zhu},
  \citenamefont {Lao}, \citenamefont {Ho}, \citenamefont {Campbell},\ and\
  \citenamefont {Hudson}}]{zhu_high-resolution_2022}%
  \BibitemOpen
  \bibfield  {author} {\bibinfo {author} {\bibfnamefont {G.-Z.}\ \bibnamefont
  {Zhu}}, \bibinfo {author} {\bibfnamefont {G.}~\bibnamefont {Lao}}, \bibinfo
  {author} {\bibfnamefont {C.}~\bibnamefont {Ho}}, \bibinfo {author}
  {\bibfnamefont {W.~C.}\ \bibnamefont {Campbell}},\ and\ \bibinfo {author}
  {\bibfnamefont {E.~R.}\ \bibnamefont {Hudson}},\ }\bibfield  {title}
  {\bibinfo {title} {High-resolution laser-induced fluorescence spectroscopy of
  ${^{28}\text{Si}^{160}}\text{O}^+$ and ${^{29}\text{Si}^{160}}\text{O}^+$ in
  a cryogenic buffer-gas cell},\ }\href
  {https://doi.org/10.1016/j.jms.2022.111582} {\bibfield  {journal} {\bibinfo
  {journal} {Journal of Molecular Spectroscopy}\ }\textbf {\bibinfo {volume}
  {384}},\ \bibinfo {pages} {111582} (\bibinfo {year} {2022})}\BibitemShut
  {NoStop}%
\bibitem [{\citenamefont {Hudson}\ and\ \citenamefont
  {Campbell}(2021)}]{hudson_laserless_2021}%
  \BibitemOpen
  \bibfield  {author} {\bibinfo {author} {\bibfnamefont {E.~R.}\ \bibnamefont
  {Hudson}}\ and\ \bibinfo {author} {\bibfnamefont {W.~C.}\ \bibnamefont
  {Campbell}},\ }\bibfield  {title} {\bibinfo {title} {Laserless quantum gates
  for electric dipoles in thermal motion},\ }\href
  {https://doi.org/10.1103/PhysRevA.104.042605} {\bibfield  {journal} {\bibinfo
   {journal} {Physical Review A}\ }\textbf {\bibinfo {volume} {104}},\ \bibinfo
  {pages} {042605} (\bibinfo {year} {2021})}\BibitemShut {NoStop}%
\bibitem [{\citenamefont {Chou}\ \emph {et~al.}(2017)\citenamefont {Chou},
  \citenamefont {Kurz}, \citenamefont {Hume}, \citenamefont {Plessow},
  \citenamefont {Leibrandt},\ and\ \citenamefont
  {Leibfried}}]{chou_preparation_2017}%
  \BibitemOpen
  \bibfield  {author} {\bibinfo {author} {\bibfnamefont {C.~W.}\ \bibnamefont
  {Chou}}, \bibinfo {author} {\bibfnamefont {C.}~\bibnamefont {Kurz}}, \bibinfo
  {author} {\bibfnamefont {D.~B.}\ \bibnamefont {Hume}}, \bibinfo {author}
  {\bibfnamefont {P.~N.}\ \bibnamefont {Plessow}}, \bibinfo {author}
  {\bibfnamefont {D.~R.}\ \bibnamefont {Leibrandt}},\ and\ \bibinfo {author}
  {\bibfnamefont {D.}~\bibnamefont {Leibfried}},\ }\bibfield  {title} {\bibinfo
  {title} {Preparation and coherent manipulation of pure quantum states of a
  single molecular ion},\ }\href {https://doi.org/10.1038/nature22338}
  {\bibfield  {journal} {\bibinfo  {journal} {Nature}\ }\textbf {\bibinfo
  {volume} {545}},\ \bibinfo {pages} {203} (\bibinfo {year}
  {2017})}\BibitemShut {NoStop}%
\bibitem [{\citenamefont {Furey}\ \emph {et~al.}(2024)\citenamefont {Furey},
  \citenamefont {Wu}, \citenamefont {Isaza-Monsalve}, \citenamefont {Walser},
  \citenamefont {Mattivi}, \citenamefont {Nardi},\ and\ \citenamefont
  {Schindler}}]{furey2024strategies}%
  \BibitemOpen
  \bibfield  {author} {\bibinfo {author} {\bibfnamefont {B.~J.}\ \bibnamefont
  {Furey}}, \bibinfo {author} {\bibfnamefont {Z.}~\bibnamefont {Wu}}, \bibinfo
  {author} {\bibfnamefont {M.}~\bibnamefont {Isaza-Monsalve}}, \bibinfo
  {author} {\bibfnamefont {S.}~\bibnamefont {Walser}}, \bibinfo {author}
  {\bibfnamefont {E.}~\bibnamefont {Mattivi}}, \bibinfo {author} {\bibfnamefont
  {R.}~\bibnamefont {Nardi}},\ and\ \bibinfo {author} {\bibfnamefont
  {P.}~\bibnamefont {Schindler}},\ }\bibfield  {title} {\bibinfo {title}
  {Strategies for implementing quantum error correction in molecular
  rotation},\ }\href@noop {} {\bibfield  {journal} {\bibinfo  {journal} {arXiv
  preprint arXiv:2405.02236}\ } (\bibinfo {year} {2024})}\BibitemShut {NoStop}%
\end{thebibliography}%

\end{document}